\documentclass[preprint]{aastex}
\newcommand{\ltsim}{\raise 2pt \hbox {$<$} \kern-1.1em \lower 4pt \hbox {$\sim$}}
\newcommand{\gtsim}{\raise 2pt \hbox {$>$} \kern-1.1em \lower 4pt \hbox {$\sim$}}

\slugcomment{ApJ vol. 551 - Apr 20, 2001 issue, in press}
\shortauthors{Giovannini, Cotton, Feretti et al.}
\shorttitle{A Complete Sample of Radio Galaxies}

\begin{document}

\title{VLBI Observations of a Complete Sample of Radio Galaxies.\\
       10 Years Later}

\author{G. Giovannini\altaffilmark{1,2}, 
W.D. Cotton\altaffilmark{3}, L. Feretti\altaffilmark{2}, 
L. Lara\altaffilmark{4}, and T. Venturi\altaffilmark{2}} 

\altaffiltext{1}{Dipartimento di Fisica, Universita' di Bologna,via B.Pichat 
6/2, 40127 Bologna, Italy}
\altaffiltext{2}{Istituto di Radioastronomia del CNR, via Gobetti 101, 40129 
Bologna, Italy}
\altaffiltext{3}{National Radio Astronomy Observatory, 520 Edgemont Rd, 
Charlottesville \\
VA 22903-2475, USA}
\altaffiltext{4}{Instituto de Astrof\'{\i}sica de Andaluc\'{\i}a, CSIC, Apdo 3004, 
18080
Granada, Spain}
\email{ggiovann@ira.bo.cnr.it,bcotton@nrao.edu,lferetti@ira.bo.cnr.it,
lucas@iaa.es,tventuri@ira.bo.cnr.it}

\begin{abstract}
A complete sample of 27 radio galaxies was selected from the B2 and 3CR 
catalogs, in order to study their properties on the milliarcsecond scale.
In the Appendix of this paper we present new radio images for 12 of them.
Thanks to the present data, all the sources in this sample have been 
imaged at mas resolution.
We discuss the general results.
In particular we stress the evidence for high velocity jets in low power
radio galaxies, we compare high and low power sources, and discuss the source 
properties in the 
light of the unified scheme models. We derive that the properties of
parsec scale jets are similar in sources with different total radio
power and kpc scale morphology.

From the core - total radio power correlation, we estimate that relativistic 
jets with Lorentz factor $\gamma$ in the range 3 - 10 are present in high 
and low power radio sources.
We discuss also the possible existence of a two velocity structure in parsec scale jets (fast
spine and lower velocity external shear layer).

\end{abstract}


\vfill
\eject

\section{Introduction}

The study of the parsec scale properties of radio galaxies is crucial to obtain
information on the nature of their central engine, and provides the basis of
the current {\it unified theories} (see e.g. \citep{ur95}), which suggest that 
the appearance of active
galactic nuclei strongly depends on orientation. In the {\it high-luminosity
unified scheme}, quasars and powerful radio galaxies (FR II, \citep{fa74}) 
are suggested to
be the same class of objects but seen at different viewing angles. 
Similarly, the
{\it low-luminosity unified scheme} assumes BL-Lacs to be the beamed population
of radio galaxies of low-intermediate luminosity (FR I).

To get new insight in the study of radio galaxies at pc resolution,
we undertook a project of observations of a complete sample of radio
galaxies selected from the B2 and 3CR catalogs \citep{g90}.
Since both catalogs have been selected at low frequencies, the source
properties are dominated by the unbeamed extended emission
and are not affected by observational biases related to orientation effects.

In the Appendix we present new images and data for 12 galaxies of our
sample and give short notes reviewing also published results. 
After 10 years high quality images and data
are available for all sources in  the sample, so that 
we can derive and discuss general properties of the whole sample.

We will use here a Hubble constant H$_0$ = 50 km sec$^{-1}$ Mpc$^{-1}$ and
a deceleration parameter q$_0$ = 0.5
(note that different cosmological constants have been used in previous 
papers on the same sample).

\section{The Sample}

In Table 1 we give the list of radio galaxies. The complete
sample \citep{g90}, consisting of 27 radio galaxies from the B2 and 3CR 
catalogs, satisfies the following criteria:

Declination $>$ 10$^\circ$

Galactic latitude b $>$ 15 $^\circ$

Arc-second core flux density at 5 GHz (S$_c$) ~$\ge$ 100 mJy

Apparent visual magnitude of the associated galaxy m$_v$ $\le$ 20, for 3CR 
sources.
 
With respect to the original list presented in \citet{g90}, we give a more 
accurate flux density measurement of the core at arc-second 
resolution, when 
available, to minimize the contamination from
the jet(s) flux density. The total radio power has been scaled according
to the different cosmological parameters used here.

\begin{deluxetable}{cccllcc}
\tabletypesize{\small} 
\tablecaption{The Selected Sample\label{tab1}}
\tablehead{
\colhead{Name}&\colhead{Name}&\colhead{z}&\colhead{S$_c$(5.0)}&
\colhead{Log P$_t$}&\colhead{Type}&\colhead{References}\\
\colhead{IAU}&\colhead{other}&\colhead{ }&\colhead{mJy}&\colhead{W/Hz}&
\colhead{ }&\colhead{ }}
\startdata
0055+30  & NGC315  & ~~0.0167 &  588       & 24.56     & FR-I &    1      \\ 
0104+32  & 3C31    & ~~0.0169 &   92       & 25.11     & FR-I &    2      \\
0116+31  & 4C31.04 & ~~0.0592 &   32       & 25.71     & CSO  &    3 * \\
0206+35  & 4C35.03 & ~~0.0375 & 106        & 25.46     & FR-I &    2   \\  
0220+43  & 3C66B   & ~~0.0215 & 182        & 25.59     & FR-I &    *      \\ 
0222+36  &         & ~~0.0327 & 122        & 24.20     & LPC  &    *     \\ 
0258+35  & NGC1167 & ~~0.0160 &\ltsim 243  & 24.65     & CSS  &    *     \\ 
0331+39  & 4C39.12 & ~~0.0202 &  149       & 24.49     & LPC  &    *     \\ 
0410+11  & 3C109.0 & ~~0.3056 &   244      & 27.78     & FR-II&    4  \\ 
0648+27  &         & ~~0.0409 &   213      & 24.31     & LPC  &    *    \\ 
0755+37 & NGC2484  & ~~0.0413 &   195      & 25.65     & FR-I &    4   \\ 
0836+29 & 4C29.30  & ~~0.0790 &   152      & 25.70     & FR-I &    5  \\ 
1101+38  & Mkn 421 & ~~0.0300 &   640      & 24.66     & BL-Lac &  6  \\ 
1142+20 &  3C264   & ~~0.0206 &   200      & 25.46     & FR-I &    2   \\ 
1144+35 &          & ~~0.0630 &   450      & 24.95     & FR-I &    7  \\ 
1217+29 &NGC4278   & ~~0.0021 &    63      & 21.44     & LPC  &    *      \\ 
1222+13 & 3C272.1  & ~~0.0037 &   180      & 23.27     & FR-I &    *    \\ 
1228+12 & 3C274    & ~~0.0037 &  4000      & 25.07     & FR-I &    8  \\ 
1322+36 & NGC5141  & ~~0.0175 &   150      & 24.36     & FR-I  &   *    \\ 
1441+52 &  3C303   & ~~0.1410 &   181      & 26.74     & FR-II &   *    \\ 
1626+39 & 3C338    & ~~0.0303 &   105      & 25.86     & FR-I &    9    \\ 
1641+17 &  3C346   & ~~0.1620 &   220      & 26.98     & FR-II &   3   \\ 
1652+39 & Mkn 501  & ~~0.0337 &   1250     & 24.96     & BL-Lac &  6   \\
1833+32 & 3C382    & ~~0.0586 &    188     & 26.31     & FR-II &   4 *   \\
1845+79 & 3C390.3  & ~~0.0569 &    340     & 26.58     & FR-II &   10  \\
2243+39 & 3C452.0  & ~~0.0811 &    130     & 26.92     & FR-II &   *      \\
2335+26 & 3C465    & ~~0.0301 &    246     & 25.91     & FR-I &    5 \\
\enddata
\tablecomments{ S$_c$(5.0) is the arc-second core flux density at 5.0 GHz; 
Log P$_t$ is the logarithmic of the total radio power at 408 MHz. 
{\bf References}: main reference to published VLBI data: 1 - \citet{cot99}; 
2 - \citet{lar97}; 3 - \citet{cot95}; 4 - 
\citet{g94}; 5 - \citet{ven95}; 6 - \citet{g99b}; 7 - \citet{g99a}; 8 - 
\citet{jun99},
and \citet{bir99}; 9 - \citet{g98a}; 10 - \citet{al96}; an * is given when 
new data are presented in this paper}
\end{deluxetable}

\begin{figure}
\epsscale{0.9}
\plotone{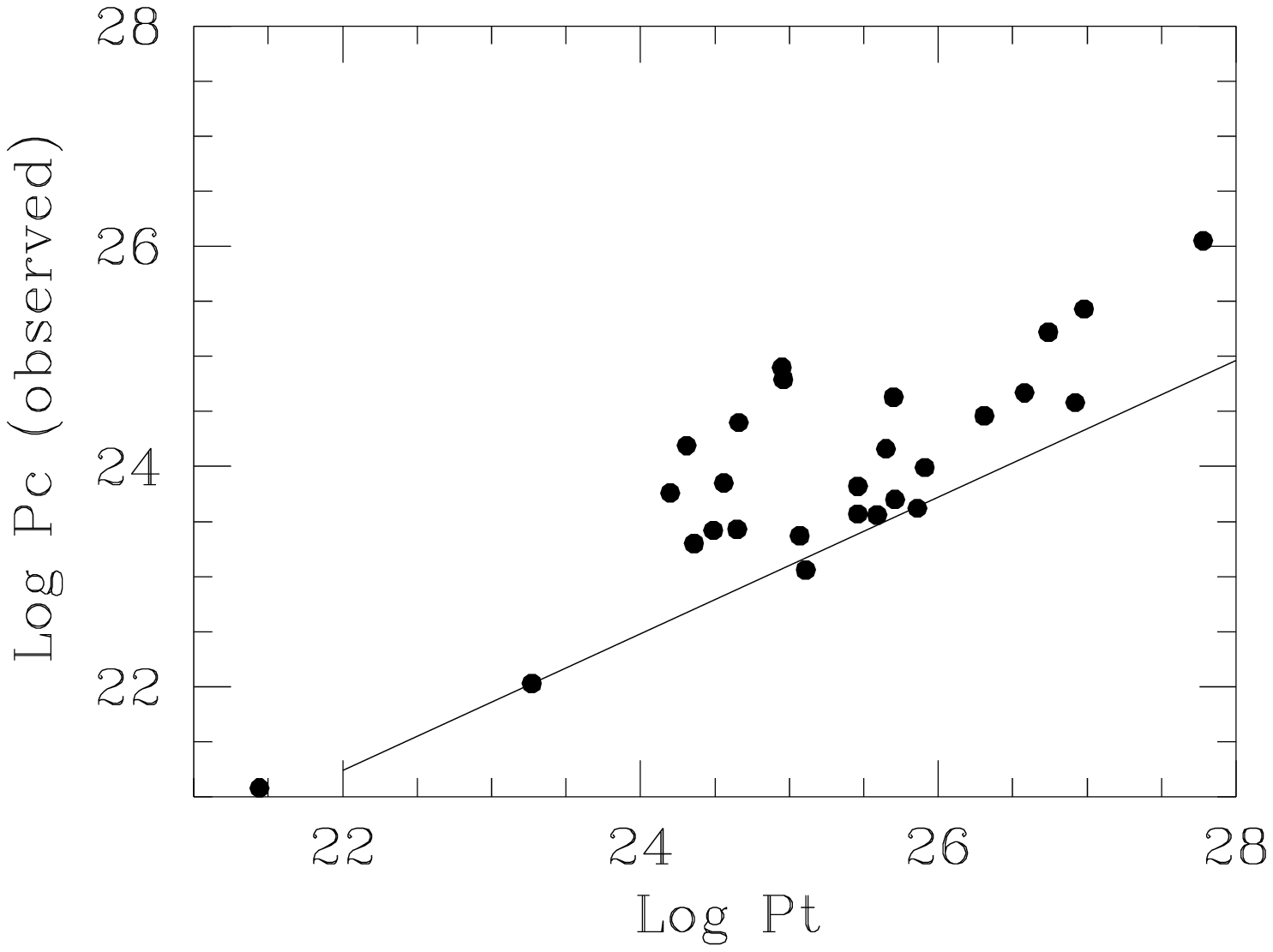}
\caption{Total radio power at 408 MHz versus the observed arc-second core radio
power at 5 GHz. The continuum line represents the correlation between the
core and total radio power found by \citet{g88}.}
\label{f1eps}
\end{figure}

The selection criterion of restricting the sample to the radio galaxies
with a relatively high core flux density limit (\gtsim 100 mJy at 6 cm),
was adopted for obvious observational reasons at the time of the sample 
selection. We are aware that it 
results in a selection effect in favor of objects with a beamed core
i.e. oriented at a small angle
with respect to the line of sight. Despite this effect,
the sample is dominated by low-intermediate power FR I radio galaxies and
by extended steep spectrum FR II radio galaxies. 
The VLBI study of a sample selected without any restrictions
on the core flux density is in progress.

In Fig. 1 we show a plot of the observed core power versus total radio power 
for the present 
sample. The line represents the correlation found by \citet{g88}.
As expected from the core flux density selection effect,
most of present sources have a core radio power higher than average with 
respect to their total radio power.

\section{Constraints on the jet orientation and velocity}

We will use the available observational data
to constrain the jet velocity and orientation for all the sources of our sample.
In the following we briefly summarize the methods used. 
Other methods are possible e.g. 
 the X-ray non thermal nuclear emission could be used to derive a possible 
Doppler
factor according to the Synchrotron-Self-Compton model (see e.g. \citet{g94}). 
However,
they do not give significant limits because of the large uncertainties
in the available data.

\subsection{Jet Sidedness}

Assuming that the jets are intrinsically symmetric, the jet
velocity ($\beta$c) and orientation ($\theta$), can be constrained from the jet to counter jet
brightness ratio R, according to the formula

\centerline {R = (1 + $\beta$ cos$\theta$)$^{2+\alpha}$ (1 - $\beta$ 
                  cos$\theta$)$^{-(2+\alpha)}$}

We assume a jet spectral index $\alpha$ = 0.5 with S$\propto\nu^{-\alpha}$. 
The validity of this formula depends on the degree of isotropy of the intrinsic
synchrotron emissivity in the jets. Here we assume that the parsec scale jet
emissivity is isotropic (see \citet{g94} for a
more detailed discussion). 

\subsection{Core Dominance}

\citet{g88} found a general correlation between the core and total radio power
in radio galaxies. Since the total radio power
was measured at low frequency and is therefore not affected by Doppler
boosting, the core emission derived by the total radio power is not boosted.
Assuming that sources are oriented at random angles the best fit value 
corresponds to the average orientation angle (60$^\circ$) and
the observed dispersion of the core radio power around the best fit line 
reflects the different orientation angles (see \citet{g94} for a more detailed
discussion).
We can use this correlation to derive the expected intrinsic
core radio power from the total galaxy radio power and, comparing it with
the observed core radio power, to estimate the source orientation.
 
Here we have re-analyzed the 
correlation given in \citet{g88} to take into account new better quality 
data of the core
flux density at 5 GHz. 
Moreover, we added the radio quasars with the same selection
criteria used for radio galaxies, since all
sources with $\theta$ in the range 0$^\circ$ - 90$^\circ$ should be included,
to derive the correct correlation.
The new correlation (scaled for the different cosmological constants used here)
is very similar to the previous one:

\centerline {log P$_c$ = (0.62$\pm$0.04) log P$_t$ + (7.6 $\pm$ 1.1)}

where P$_c$ is the arc-second core radio power at 5 GHz and P$_t$ is the
total radio power at 408 MHz. To take into account the core variability,
we have allowed the core flux density to vary within a factor of two from 
the measured one.

\subsection {Proper Motion}

Proper motion is detected in 7 sources, for which we can derive
the apparent velocity $\beta_a$~c. From this value we can constrain the intrinsic
jet velocity and orientation as follows:

\centerline {$\beta$ = $\beta_a$/($\beta_a$ cos $\theta$ + sin $\theta$)}.

In this paper we will assume that the jet bulk velocity and the pattern 
velocity have
comparable values (see e.g. \citet{ghi93}).

\section {Results}

In Table 2 and Figures 2 and 3, we give for each source the estimated range 
of the jet velocity 
and orientation with respect to the line of sight derived from the methods
previously discussed.

\begin{deluxetable}{ccccccc}
\tabletypesize{\small} 
\tablecaption{Jet velocity and orientation - estimated parameters\label{tab2}}
\tablehead{
\colhead{Name}&\colhead{Name}&\colhead{Type}&\colhead{$\theta$ range}&
\colhead{$\beta$ range}&\colhead{$\gamma$ range}&\colhead{Motion}\\
\colhead{IAU}&\colhead{other}&\colhead{ }&\colhead{degree}&\colhead{v/c}&
\colhead{ }&\colhead{$\beta_a$}}
\startdata
0055+30 & NGC 315 & FR I & 30 - 40  & $>$ 0.8       & $>$ 1.7  & 1.13 - 2.51 \\
0104+32 & 3C 31   & FR I & 40 - 60  & $>$ 0.7       & $>$ 1.4  &             \\
0116+31 & 4C31.04 & CSO  & $>$ 75   & any           &   --     &             \\
0206+35 & 4C35.03 & FR I & $<$ 54   & $>$ 0.5       & $>$ 1.15 &             \\
0220+43 & 3C66B   & FR I & $\sim$ 45& 0.6 - 0.99 & 1.25 - 7.09 &            \\
0222+36 &         & LPC  & $<$ 40   & $>$ 0.7       & $>$ 1.4  &            \\
0258+35 & NGC 1167& CSS  &  --      &   --          &  --      &            \\
0331+39 & 4C39.12 & LPC  & $<$ 45  & $>$ 0.5       & $>$ 1.15 &            \\
0410+11 & 3C 109  & FR II& 10 - 35  & $>$ 0.8       & $>$ 1.7  &             \\
0648+27 &         & LPC  & $<$ 40   & $>$ 0.7       & $>$ 1.4  &             \\
0755+37 & NGC 2484& FR I & $<$ 45   & $>$ 0.6       & $>$ 1.25 &            \\
0836+29 & 4C29.30 & FR I & $<$ 35   & $>$ 0.55      & $>$ 1.20 &            \\
1101+38 & Mkn 421 & BL-Lac& $<$ 30  & $>$ 0.87      & $>$ 2.03 & 1.5        \\
1142+20 & 3C 264  & FR I & $\sim$ 50& $\sim$ 0.98   & $\sim$ 5.0 &          \\
1144+35 &         & FR I & 20 - 25  & $\ge$ 0.95    & $\ge$ 3.2&   2.7     \\
1217+29 & NGC 4278& LPC  & --       &  --           &  --      &           \\
1222+13 & 3C 272.1& FR I & 60 - 65  & \gtsim 0.9  & $\ge$ 2.29 &         \\
1228+12 & 3C 274  & FR I & $<$ 19   & \gtsim 0.99 & \gtsim 6 &  6      \\
1322+36 & NGC 5141& FR I &\ltsim 58& \gtsim 0.54& \gtsim 1.19&       \\
1441+52 & 3C 303  & FR II& $\le$ 40 & $\ge$ 0.7     & $>$ 1.4  &          \\
1626+39 & 3C 338  & FR I & $\sim$ 85& $\sim$ 0.8    & $\sim$ 1.7& 0.8 - 0.9 \\
1641+17 & 3C 346  & FR II& $<$ 30   & $>$ 0.8       & $>$ 1.7  &          \\
1652+39 & Mkn 501 & BL-Lac& 10 - 15 & $>$ 0.99 - 0.999& 7.09 - 22.4& 4 - 8   \\
1833+32 & 3C 382  & FR II & $<$ 45  & $>$ 0.6       & $>$ 1.25 &           \\
1845+79 & 3C 390.3& FR II & 30 - 35 & $>$ 0.96       & $>$ 2.29 &  3.5     \\
2243+39 & 3C 452  & FR II &\gtsim 60 & $>$ 0.4    & $>$ 1.09 &          \\
2335+26 & 3C 465 & FR I & \ltsim 54 & $>$ 0.6     & $>$ 1.25 &         \\
\enddata
\tablecomments{$\theta$ gives the allowed values for the jet orientation
with respect to the line of sight. $\beta$ and $\gamma$ gives the 
allowed values
for the jet velocity. $\beta_a$~c is the measure of the apparent velocity, 
where available.}
\end{deluxetable}

\begin{figure}
\epsscale{0.8}
\plotone{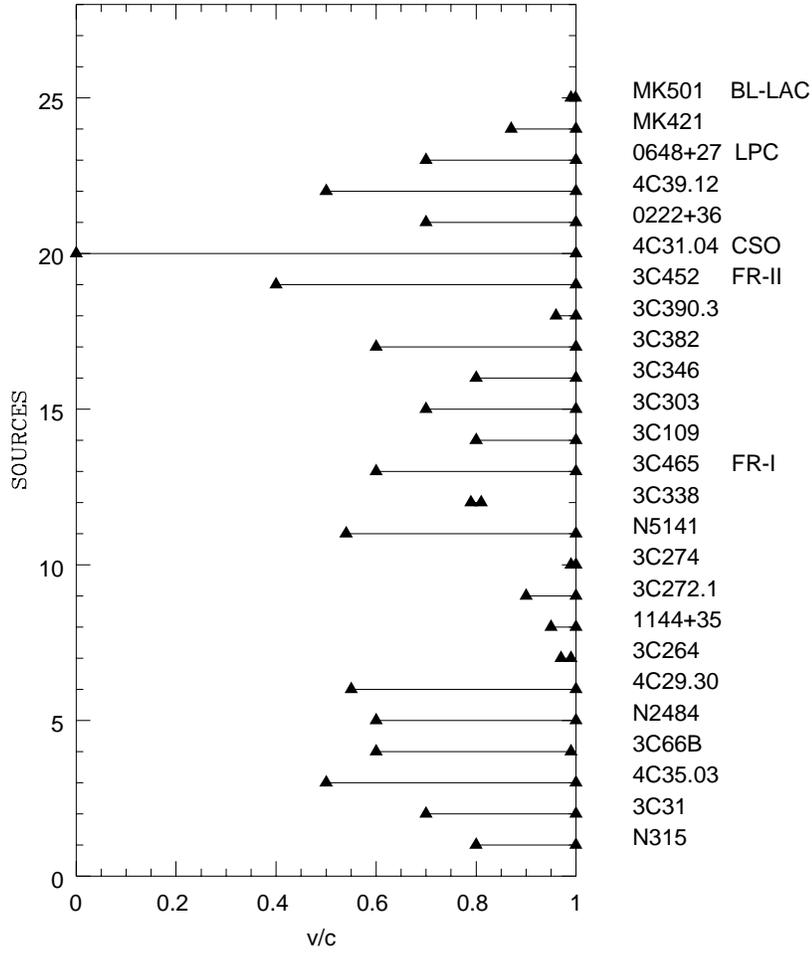}
\caption{Allowed range of the jet velocity $\beta$ = v/c 
for different sources. Note that sources are grouped together 
according to their large scale radio morphology.}
\label{f2eps}
\end{figure}

\begin{figure}
\epsscale{0.8}
\plotone{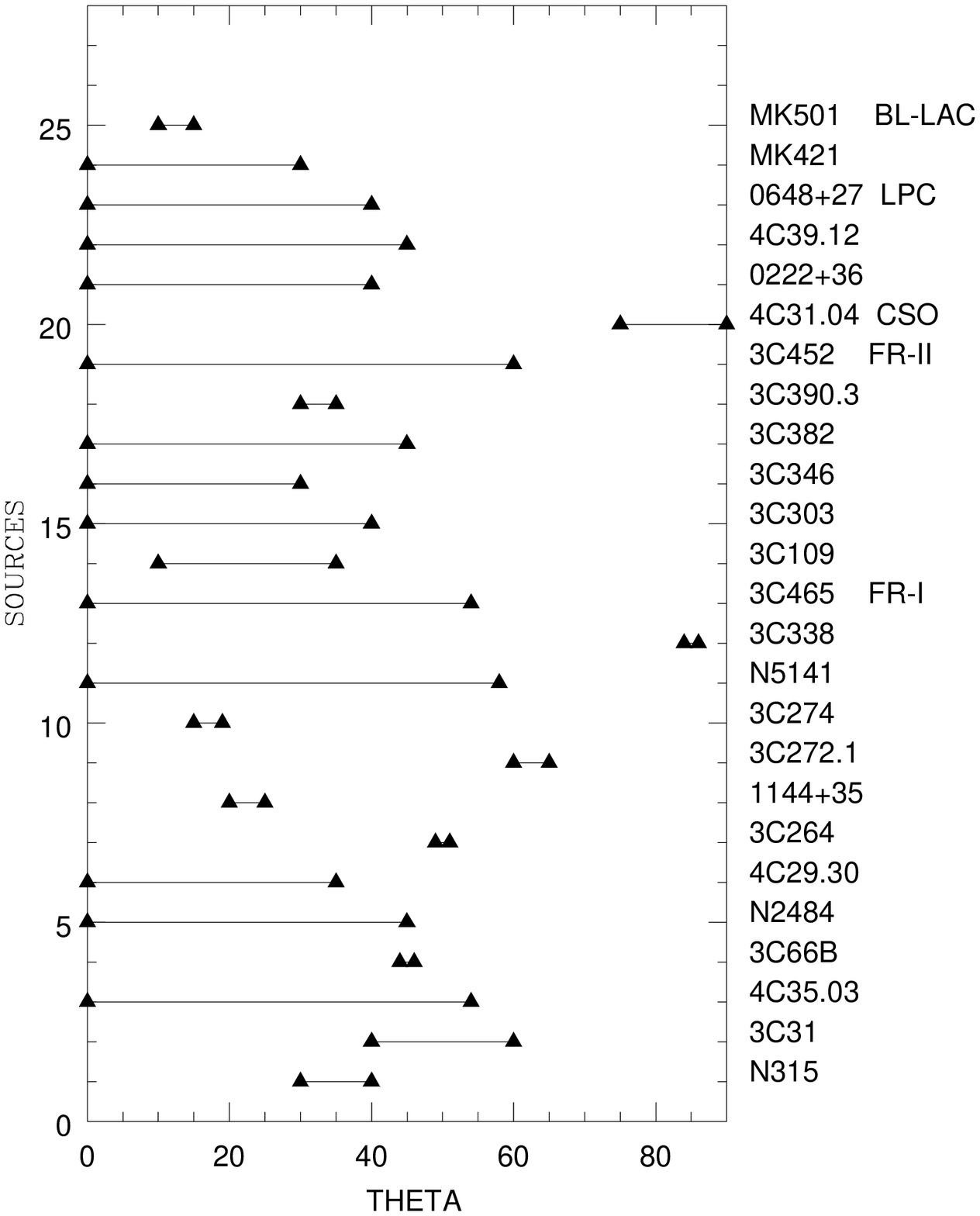}
\caption{Allowed range in the orientation angle $\theta$ with respect to the line
of sight for different sources. Note that sources are grouped together 
according to their large scale radio morphology.}
\label{f3eps}
\end{figure}

In most sources the allowed ranges in jet velocity and/or orientation are 
large because of: i) the low brightness of the jets, ii)
the uncertainties in the core dominance related to the possible core flux 
density variability, iii) the lack of measured proper motion.
For 2 sources of our sample (0258+35 and 1217+29), we cannot give any 
constraint, therefore we have a working sample of 25 sources.

\subsection {FR I Radio Galaxies}

In the present sample, there are 13 FR I radio galaxies: one (3C338) shows a 
clear two-sided structure,
three show a short counter-jet (NGC 315, 3C66B, and 1144+35), and nine  have 
a one-sided morphology. 

In all the FR I sources, the jets are 
relativistic, $\beta$ being always larger than 0.5 and in four
cases $>$ 0.9 (Fig. 2). 
Proper motion has been measured in four sources (Table 2),
and could probably be found in other sources but it requires a larger
observational effort. The jet orientation is in agreement
with the expectations of the unified model: $\theta$ is $>$
30$^\circ$ for six sources while in five sources we do not have strong
constraints on  $\theta$ which could be  $>$ 30$^\circ$, but
smaller values cannot be excluded.
In 3C~274 and 1144+35
a viewing angle smaller than 30$^\circ$ is obtained (Fig. 3). 
However, 1144+35 shows intermediate properties
between FR I sources and BL-Lac type objects \citep{g99a} 
and 3C 274 
is a peculiar source (see Sect. 4.4).

Two sources (NGC 315 and 3C 66B) show evidence of an accelerating jet near
the core (1 - 5 parsec; see Appendix).

Parsec scale jets are always collimated near the core (with the exception
of 3C\ 274, \citep{jun99}) and appear rather uniform in brightness, with the 
presence of possible jet substructures at any distance from the core.
In 3C\ 274 it is well known that the jet appears
limb-brightened; a similar structure is visible in 1144+35 at about 20 pc 
from the core (projected).
We cannot exclude that a limb-brightened morphology could be present in 
a larger number of
sources, but higher sensitivity and higher resolution data are 
necessary to reveal it (see Sect. 4.5). 

Furthermore, we point out that we are imaging different regions of the parsec 
scale jets in different sources,  
given the different redshift and orientation angle of the
present sources.
Among the 13 FR I radio galaxies, only in three cases we are able to
image the jet out to more than 50 pc from the core (de-projected distance):
3C 264, 1144+35, and 3C\ 274. In 1144+35 and 3C\ 274 we find evidence of a 
limb-brightened jet structure in this distant region; in 3C 264 the jet 
is well collimated and narrow up to $\sim$ 100 pc and it shows a strong 
widening, filaments and a possible helical structure at a 
larger distance from the core. 
At a shorter distance from
the core (1 to 50 pc) we have in all cases centrally peaked jets, except 
3C\ 274.
In a few cases (see e.g. 3C~66B) there is a sharp change in
the jet brightness, but no critical scale was found. Only for a few sources
can we obtain images on linear scales smaller than 1 pc. 
In these cases, we 
see a collimated, centrally peaked jet with the exception of 3C\ 274. 
In this source, an extended jet emission is visible, possibly 
related to the collimation process. A strong collimation of the jet occurs on 
apparent scales \gtsim 0.04 pc from the central engine
(see \citet{jun99}). 

\subsection {FR II Radio Galaxies}

Among the FR II sources, one shows a symmetric two-sided structure (3C452)
while five have a one-sided jet (3C109, 3C303, 3C346, 3C382, and 3C390.3). This 
result is in very good agreement with the unified scheme model 
predictions, because
3C452 the only Narrow Line FR II galaxy in our sample, while other FR II 
galaxies are well known Broad Line Radio Galaxies. 
The main jet in the
pc scale image is always on the same side with respect to the
core as the main kpc scale jet, and the jet position angles are in agreement 
within a few degrees.
This is expected if the jet asymmetry present in FR II radio galaxies
is due to the Doppler effect in jets moving at
relativistic velocities in the pc and kpc scale.   

Comparing parsec scale jets in FR I and FR II sources we note that in our
images FR I jets show a uniform brightness when observed with a good 
uv-coverage, while FR II jets show a structure characterized by the presence
of a few blobs. This difference is difficult to quantify and could be
related to a difference in the jet properties. We exclude that it could be 
an effect due to the {\it clean} procedure and poor uv-coverage,
since the three FR II images
presented here for the first time were obtained with an array of 17
telescopes (see Appendix).

No evidence of
a larger jet velocity in FR II than in FR I sources is present in our data
(see Fig. 2).

\subsection {Compact Sources}

The compact sources there include two BL-Lac type objects (Mkn 421 and 
Mkn 501), four Low Power Compact Sources (LPC: 0222+36, 0331+39, 0648+27, and
1217+29), one Compact Symmetric Object (0116+31) and one Compact Steep Spectrum
source (0258+35).

The two BL-Lacs show
high velocity pc-scale jets oriented at a small angle with
respect to the line of sight. In Mkn 421 we see a complex jet with no
evidence of a limb-brightened structure whereas in Mkn 501 a limb-brightened 
jet is well visible at $\sim$ 50 
pc from the core (de-projected).

In two LPC sources there is evidence
of fast jets oriented at a small angle with respect to the line of sight.
In one case (0331+39) the jet is visible for more than 50 pc and could be
limb-brightened starting at about 15 pc from the core (de-projected). In
 0222+36 no jet was detected, but the core dominance
suggests a source orientation at a small angle with respect to the line
of sight and some relativistic boosting due to a relativistic jet not visible
in our images.
The evidence of relativistic jets oriented at a small angle with respect 
to the 
line of sight and the absence of an extended structure at 
arc-second resolution,
suggests that these sources could be classified as intermediate or low power
BL-Lac sources whose observed core power is too low
to dominate the optical emission and therefore are classified as galaxies.
We remember that LPC sources do not have a steep radio spectrum.

NGC 4278 shows diffuse emission on a linear scale
smaller than 1 pc. All the radio emission is confined within
this small size source. No jet like structure is clearly visible and no 
constraint can be given on the source orientation.
More observations are necessary to properly discuss
this source.

For the CSO and CSS source we refer to the comments given in the Appendix.

\subsection {High Velocity Jets}

There are several strong and widely accepted lines of evidences for the 
existence of relativistic bulk velocities in the 
parsec scale jets of radio galaxies: the observed super-luminal motions, the
rapid variabilities, the observed high brightness temperatures, the absence 
of strong inverse-Compton emission in the X-ray and the 
detection of a high frequency emission (gamma ray) for the
two BL-Lacs Mkn 421 and 501 all seem to require relativistic bulk speeds with
Lorentz factor ($\gamma$) \gtsim 3.

The results, discussed in the previous sections, confirm that
radio jets move at high velocities on the mas scale. 
Since in many cases 
we can only give a lower limit to $\gamma$, to better investigate this point 
we assumed different 
$\gamma$ values and we tested if the derived source properties 
were in agreement with the observational data.
Once a jet velocity is assumed, the jet orientation is constrained by the
observational data and it is possible to compute the corresponding Doppler 
factor $\delta$ ($\delta$ = ($\gamma (1 - \beta cos \theta))^{-1}$ for each
source (in Table 3 we present the estimated orientation angle and Doppler
factor ranges assuming $\gamma$ = 5). Then,
from the value of $\delta$ and of the measured
radio power, we can derive the intrinsic core radio power for each source:
P$_{c-observed}$ = P$_{c-intrinsic}$ $\times$ $\delta^2$ (assuming $\alpha$
= 0). In Table 4 we report the intrinsic radio power (assuning $\gamma$ = 5),
and for a comparison,
the observed total and core radio power. 
Since there is a range of possible jet orientations, we have a possible
range of values for $\delta$ and therefore of P$_{c-intrinsic}$. 

\begin{deluxetable}{ccccc}
\tabletypesize{\small} 
\tablecaption{Jet orientation and Doppler factor with $\gamma$ = 5\label{tab3}}
\tablehead{
\colhead{Name}&\colhead{Name}&\colhead{Type}&\colhead{$\theta_5$}&
\colhead{$\delta_5$}\\
\colhead{IAU}&\colhead{other}&\colhead{ }&\colhead{degree}&\colhead{ }}
\startdata
0055+30 & NGC 315 & Fr I & 30 - 40 & 1.32 - 0.80 \\ 
0104+32 & 3C 31   & FR I & 50 - 60 & 0.54 - 0.39 \\
0116+31 & 4C31.04 & CSO  & 75 - 80 & 0.27 - 0.24 \\
0206+35 & 4C35.03 & FR I & 35 - 54 & 1.01 - 0.47 \\
0220+43 & 3C66B   & FR I &   45    &     0.65    \\
0222+36 &         & LPC  & 25 - 40 & 2.13 - 0.80 \\
0258+35 & NGC 1167& CSS  &   ---   &     ---     \\
0331+39 & 4C39.12 & LPC  & 35 - 45 & 1.01 - 0.65 \\ 
0410+11 & 3C 109  & FR II& 23 - 34 & 2.04 - 1.07 \\
0648+27 &         & LPC  & 30 - 40 & 1.32 - 0.80 \\
0755+37 & NGC 2484& FR I & 30 - 45 & 1.32 - 0.65 \\
0836+29 & 4C29.30 & FR I & 25 - 35 & 2.13 - 1.01 \\
1101+38 & Mkn 421 & BL-Lac&20 - 28 & 2.52 - 1.48 \\
1142+20 & 3C 264  & FR I&$\sim$ 50 &     0.54    \\  
1144+35 &         & FR I&$\sim$ 25 &     1.79    \\
1217+29 & NGC 4278& LPC  &   ---   &    ----     \\
1222+13 & 3C 272.1& FR I & 60 - 65 & 0.39 - 0.34 \\
1228+12 & 3C 274  & FR I &  9 - 10 &      6      \\
1322+36 & NGC 5141& FR I & 45 - 58 & 0.65 - 0.42 \\
1441+52 & 3C 303  & FR II& 25 - 40 & 1.79 - 0.80 \\
1626+39 & 3C 338  & FR I &   85    &    0.22     \\
1641+17 & 3C 346  & FR II& 23 - 30 & 2.04 - 1.32 \\
1652+39 & Mkn 501 & BL-Lac&10 - 15 & 5.7 - 3.73  \\
1833+32 & 3C 382  & FR II& 35 - 45 & 1.01 - 0.65 \\
1845+79 & 3C 390.3& FR II& 30 - 35 & 1.32 - 1.01 \\ 
2243+39 & 3C 452  & FR II& 60 - 70 & 0.39 - 0.30 \\
2335+26 & 3C 465  & FR I & 37 - 54 & 0.92 - 0.47 \\ 
\enddata
\tablecomments{ $\theta_5$ and $\delta_5$ are the orientation angle and Doppler
factor value assuming $\gamma$ = 5.}
\end{deluxetable}

\begin{deluxetable}{cccllcc}
\tabletypesize{\small} 
\tablecaption{Observed and Intrinsic core radio power\label{tab4}}
\tablehead{
\colhead{Name}&\colhead{Name}&\colhead{z}&\colhead{Log P$_t$}&
\colhead{Log P$_{c-observ.}$}&\colhead{Log P$_{c-intr5}$}&\colhead{Type} \\
\colhead{IAU}&\colhead{other}&\colhead{ }&\colhead{W/Hz}&\colhead{W/Hz}&
\colhead{W/Hz}&\colhead{ }}
\startdata
0055+30  & NGC315  & ~~0.0167 & 24.56      & 23.85               & 23.61 - 24.04     & FR-I   \\ 
0104+32  & 3C31    & ~~0.0169 & 25.11      & 23.06               & 23.60 - 23.88     & FR-I  \\
0116+31  & 4C31.04 & ~~0.0592 & 25.71      & 23.70               & 24.84 - 24.94     & CSO \\
0206+35  & 4C35.03 & ~~0.0375 & 25.46      & 23.82               & 23.81 - 24.48     & FR-I \\  
0220+43  & 3C66B   & ~~0.0215 & 25.59      & 23.56               & 23.93             & FR-I \\ 
0222+36  &         & ~~0.0327 & 24.20      & 23.76               & 23.10 - 23.95     & LPC   \\ 
0258+35  & NGC1167 & ~~0.0160 & 24.65      & \ltsim 23.43        &     -----         & CSS  \\ 
0331+39  & 4C39.12 & ~~0.0202 & 24.49      & 23.42               & 23.41 - 23.79     & LPC  \\ 
0410+11  & 3C109.0 & ~~0.3056 & 27.78      & 26.05               & 25.43 - 25.99     & FR-II \\ 
0648+27  &         & ~~0.0409 & 24.31      & 24.19               & 23.95 - 24.38     & LPC \\ 
0755+37 & NGC2484  & ~~0.0413 & 25.65      & 24.16               & 23.92 - 24.53     & FR-I \\ 
0836+29 & 4C29.30  & ~~0.0790 & 25.70      & 24.63               & 23.97 - 24.62     & FR-I \\ 
1101+38  & Mkn 421 & ~~0.0300 & 24.66      & 24.40               & 23.60 - 24.06     & BL-Lac \\ 
1142+20 &  3C264   & ~~0.0206 & 25.46      & 23.57               & 24.11             & FR-I \\ 
1144+35 &          & ~~0.0630 & 24.95      & 24.90               & 24.39             & FR-I \\ 
1217+29 &NGC4278   & ~~0.0021 & 21.44      & \ltsim 21.08        & -----             & LPC  \\ 
1222+13 & 3C272.1  & ~~0.0037 & 23.27      & 22.03               & 22.85 - 22.97     & FR-I \\ 
1228+12 & 3C274    & ~~0.0037 & 25.07      & 23.37               & 21.81             & FR-I \\ 
1322+36 & NGC5141  & ~~0.0175 & 24.36      & 23.30               & 23.67 - 24.05     & FR-I \\ 
1441+52 &  3C303   & ~~0.1410 & 26.74      & 25.22               & 24.71 - 25.41     & FR-II \\ 
1626+39 & 3C338    & ~~0.0303 & 25.86      & 23.62               & 24.94             & FR-I \\ 
1641+17 &  3C346   & ~~0.1620 & 26.98      & 25.43               & 24.81 - 25.19     & FR-II \\ 
1652+39 & Mkn 501  & ~~0.0337 & 24.96      & 24.79               & 23.28 - 23.65     & BL-Lac \\
1833+32 & 3C382    & ~~0.0586 & 26.31      & 24.46               & 24.45 - 24.83     & FR-II \\
1845+79 & 3C390.3  & ~~0.0569 & 26.58      & 24.67               & 24.43 - 24.66     & FR-II \\
2243+39 & 3C452.0  & ~~0.0811 & 26.92      & 24.58               & 25.63 - 25.40     & FR-II \\
2335+26 & 3C465    & ~~0.0301 & 25.91      & 23.99               & 24.06 - 24.65     & FR-I \\
\enddata
\tablecomments{Log P$_t$ is the logarithm of the total radio power at 408 MHz.
Log P$_{c-observ.}$ and Log P$_{c-intr5}$ are the logarithm of observed and intrinsic 
core radio power at 5 GHz, with $\gamma$ = 5.}
\end{deluxetable}

\begin{figure}
\epsscale{0.9}
\plotone{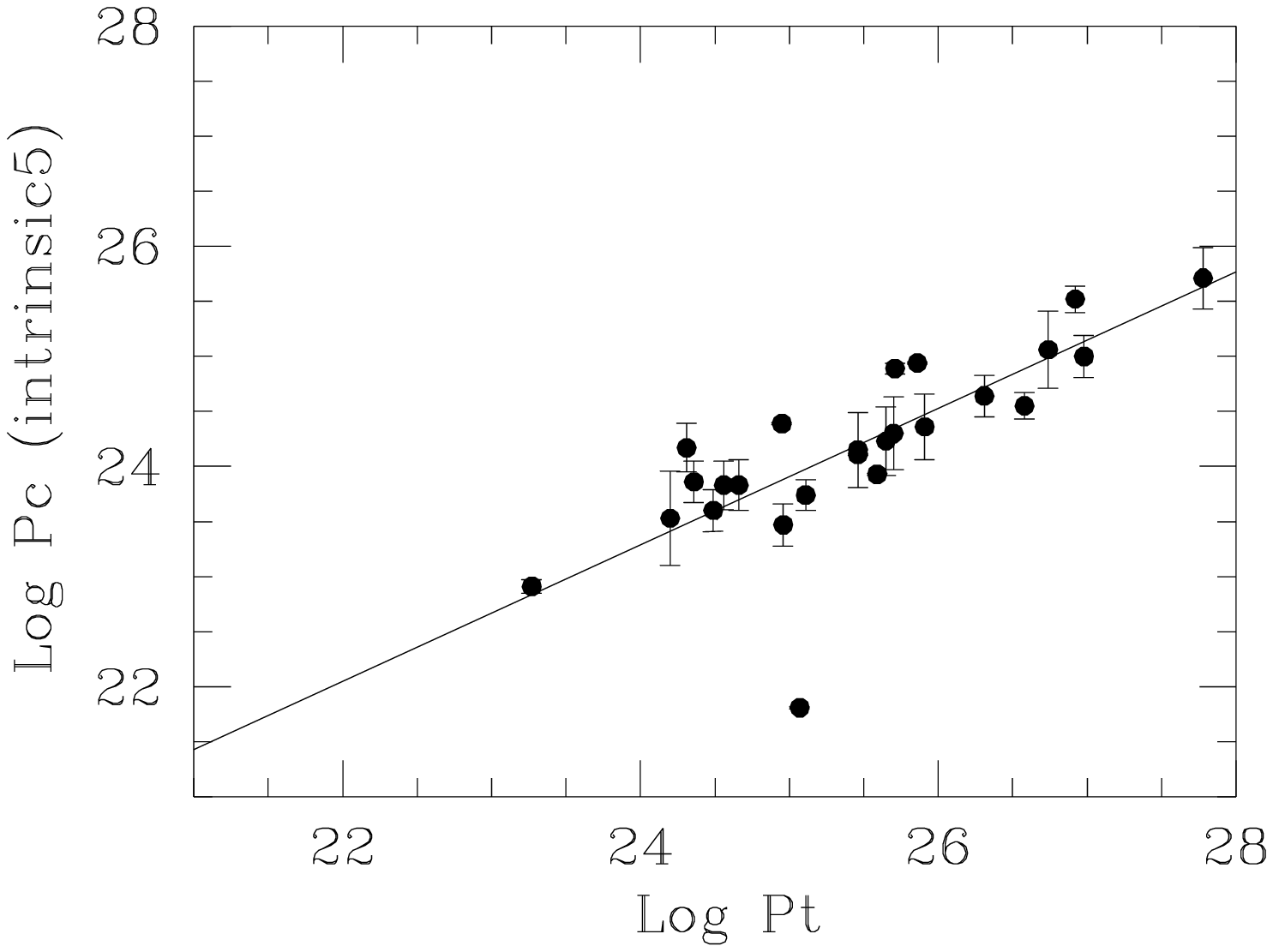}
\caption{Correlation between the core intrinsic radio power and the total radio
power. The line is not the best fit but the scaled correlation 
found between the observed
core radio power and the total radio power. $\gamma$ = 5 has been assumed.
The discrepant point corresponds to 3C274.}
\label{f4eps}
\end{figure}

We derived the intrinsic core radio power for different values of $\gamma$.
In figure 4 we show
the total core radio power at 408 MHz versus
the intrinsic core radio power at 5 GHz estimated with $\gamma$ = 5.
Vertical bars represent the possible range of the intrinsic core radio
power implied by the estimated range of $\delta$.
Despite the large range of possible values allowed in the core dominance
(see Sect. 3.2) and the possibility to have low and high power radio sources
with a different $\gamma$ value,
we find a small dispersion 
and a good agreement with the 
correlation between the observed core radio power and the total core
radio power. 
The line drawn in Fig. 4 is the same line as in Fig. 1 taking into account 
that now we are plotting the intrinsic core radio power:

\centerline {log P$_{ci5}$ = 0.62 log P$_t$ + 8.41}

where P$_{ci5}$ is the intrinsic core radio power derived assuming 
$\gamma$ = 5.
If we compare Figure 1 and Figure 4 we see that the points corresponding
to the present sample are in good agreement with the median line
and have a small dispersion around it. This is expected, since the dispersion visible
in Figure 1 (due to different orientation angles) has been removed, as 
 in Fig. 4 we plotted the intrinsic core radio power.
A similar result was obtained for $\gamma$ in the range 3 - 10.
If we assume $\gamma$ $<$ 3, the Doppler correction is such that 
the points corresponding to the present galaxies are not around the best fit
line but at a higher value (as in Fig. 1). $\gamma$ $>$ 10 
implies a higher dispersion of the observed core radio power with respect
to the point dispersion found in \citet{g88}.
Present data are in agreement with a $\gamma$ value in the
range 3 - 10.
In the future, with a larger sample
we could give stronger constraints to the value of $\gamma$.

One point in the plots is definitely too low for any $\gamma$:
the intrinsic core radio power derived for 3C 274 is always lower than 
 the value expected from its total radio power. This source would agree with 
the correlation displayed in Fig. 4 with a Doppler factor 
$\sim$ 1, which would imply a jet orientation in the range 
35$^\circ$--40$^\circ$ and not smaller than 19$^\circ$.
Such a value
is in contrast with the
high super-luminal motion found with optical data (see the discussion in 
\citet{bir99}) unless a high jet velocity ($\gamma$ $>$ 12) is assumed
(see \citet{bir99}).
Even taking into account that most of the 
total radio power in 3C 274 at low frequency is due to the large extended 
halo emission,
whose origin and relation with the core radio power is poorly understood
(see \citet{ow99} and \citet{harr99}), and that therefore the total radio 
power to be
used in the correlation should be probably much lower, the large discrepancy 
between 3C 274 and the other sources persists.
We conclude that 3C274 looks peculiar with respect to
the other sources of this sample. 

We note that all sources in our sample are in agreement with the same
P$_c$ -- P$_{tot}$ correlation, despite the variety of their large scale 
morphology (FR I, FR II, BL-Lacs, and so on). Moreover the correlation
holds over 5 orders of magnitude in P$_{tot}$ (see Fig. 4).
This result confirms that the properties of parsec scale jets 
are similar in sources with different total radio power and kpc 
scale morphology.

A correlation between the core and the total radio power can be
expected from simple considerations if the sources are in energy equipartition
conditions.
In this case the total energy (U$_l$) in the radio source lobes is:

\centerline {U$_l$ $\propto$ P$_l^{4/7}$ $\times$ V$^{3/7}$}

\citep{pa70} where V is the source volume and P$_l$ is the lobe radio power.

Assuming that the global energy is proportional to the energy carried
out by the jets and that the 
lobe radio power is the total radio power at
408 MHz, we can derive that:

\centerline {P$_c$  $\propto$ P$_t^{4/7}$ $\times$ V$^{3/7}$/t}

Where t is the radio source age.
The result, P$_c$  $\propto$ P$_t^{0.57}$, is very similar to the slope 
computed from
the core -- total radio power correlation (0.62).
The time dependence is too low to be relevant if we take into account 
that the
source volume is proportional to t$^a$ with {\it a} $\sim$ 3 resulting in
a final time dependence of t$^{2/7}$ and that
most of the sources used 
to derive the correlation between the core and total radio power are old 
sources. 
The small dispersion found and the
estimated correlation persists if sources are not too far from 
the equipartition condition.

\subsection {Kiloparsec and Parsec Scale Morphology}

The differences in the kpc scale between FR I and FR II sources are well
known and established. The most relevant difference related to the present
discussion is the evidence that kpc scale jets in FR II sources are still
relativistic or mildly relativistic up to the hot spot region, while kpc 
scale jets in FR I sources are strongly decelerated and become 
sub-relativistic within 1 - 5 kpc from the core, with typical velocities of
a few thousands km/sec.

In this paper we show that 
parsec scale jets are highly relativistic in FR I and FR II sources.
Therefore the difference between FR I and FR II sources on the kpc scale are 
not related to the parsec scale 
velocity, but to a larger scale effect. Two different scenarios
can be considered:

a) FR II sources have a higher intrinsic jet power which can maintain
relativistic jets up to the hot spot region;

b) FR I sources have a higher density interstellar medium which decelerates
the jets.

The first case implies a correlation between the total or core radio 
power of FR II sources and the jet velocity, and therefore jets in FR I should
have on average 
a lower $\gamma$ value. This is not the case from present
results, where no difference in the jet velocity has been found in FR I and
in FR II sources. Moreover the correlation between the core and total radio
power does not show any difference among FR I and FR II sources. If FR II
sources have higher velocity jets we should see some
discontinuity in the core-radio power correlation going from low to 
high powers.

If the interstellar medium (ISM) in FR I is denser than in FR II, 
we should see:
i) lower polarized flux and higher rotation measure in FR I than in FR II,
if the ISM is ionized. In NGC 315 \citep{cot99} found a very deep limit in the
polarized flux density, but data are missing for a comparison between FR I and
FR II sources. ii) FR I jets should show a more prominent limb-brightening
structure with respect to FR II sources because of a larger interaction
with the surrounding medium. As discussed in the next sub-section,
the few sources 
with an evident limb-brightened jet structure on the parsec
scale are low power (FR I) radio galaxies or BL-Lac type sources. 

In conclusion, our results do not support the idea that large scale 
differences between FR I and FR II sources are due to intrinsically 
more powerful and
faster parsec scale jets in FR II sources. There is evidence that 
the jet velocity in FR I sources slowly, but 
continuously, decreases because of the jet interaction with a dense ISM, and 
becomes sub--relativistic on the kpc scale.

\subsection {Two Velocity Jets?}

\citet{chi00}  
derive that a structure in the jet velocity field, with a fast spine  
surrounded by a slow velocity layer, is necessary to account 
for the discrepancies among
observational data for FR I radio galaxies and BL-Lac objects. 
A two-velocity jet model was developed by \citet{la96} and supported
by observational evidence in some FR I sources. In a two velocity 
jet, seen at a
significant angle to the jet, the emission from the inner portion of the jet
is strongly de-boosted, making it weak to invisible. In contrast, the outer,
slower portions of the jet have lower Doppler de-boosting and therefore may
appear brighter. Such a jet if observed at proper resolution, would appear 
as a hollow cone, i.e. limb-brightened.

In our sample
a limb-brightened structure is clearly present in 3C\ 274, 1144+35
and Mkn 501 and could be present in 0331+39. We note
that the detection of such structure may not be easy: it should be well visible
in polarized images (see \citet{aar99} for Mkn 501) which are difficult to
obtain because of the low percentage of polarized flux (see e.g. 
\citet{cot99}), while good angular 
resolution and high sensitivity are needed in 
total intensity images. Moreover to distinguish the 
two velocity
regions a critical orientation is crucial to have the low velocity external
region brighter because of Doppler beaming and the inner
fast region fainter because it is de-boosted. This is the case of 1144+35
and MKn 501. We also note that 
the two velocity structure is not visible at the 
jet beginning,
but at some distance from the core. We interpret this as evidence of the
presence of an interaction between the jet and the external medium, i.e. the 
outer layer velocity 
 decreases because of the interaction with the surrounding
medium, and this deceleration becomes significant at some distance from
the core.
The de-projected distance, where the
shear layer is visible, is about 50 pc for 1144+35 and Mkn 501, 
and 20--30 pc for 0331+39, where it could be present. 
This suggests that jet imaging in the sub-arcsecond scale (i.e. 50 -- 100 pc)
is likely to be best suited for the detection of limb-brightened structures.
Most of our images can track the milliarcsecond jets only out to a few
parsecs and only for Mkn 421 and 3C 264 high quality images from a few pc
to hundreds of parsecs are available, with no evidence for limb-brightening.
Both sources have a complex morphology
because of projection effects and jet sub-structures.
We conclude that a two--velocity structure in parsec scale jets is
present in a few sources, but with the available data we cannot confirm 
if this kind of 
structure is present in all sources or only in some of them because of 
different jet interactions with the surrounding medium. Future observations 
with the extended VLA will solve this point.

\section {Conclusions}

In this paper we presented new images at parsec resolution for 
10 galaxies (see Appendix),
and maps at different frequencies for 2 more sources. Thanks to these new data
at least one high quality image at parsec resolution is available for
all the sources of our sample.

FR I radio sources are characterized by the presence of highly relativistic
parsec scale jets, whose orientation with respect to the line of sight, 
is in agreement with the predictions of unified scheme models. 
FR II sources have a radio morphology
very similar to FR I radio galaxies and jet velocities are in the same range.
A new symmetric two-sided jet was found in the narrow line
FR II radio galaxy 3C 452. LPC sources show relativistic jets at a small
angle to the line of sight. We suggest that most of LPC sources
are intermediate BL-Lacs where the observed nuclear emission is too faint to
be dominant with respect to the parent galaxy radiation because of the
intermediate jet orientation angle and/or the faint nuclear power.

Assuming different jet velocities, we derived the corresponding intrinsic
core radio power, and we correlated it with the low frequency total radio 
power. We found that the core to total radio power ratio is in very good 
agreement
with the correlation derived from the analysis of B2 and 3CR catalogs
assuming a jet bulk speed in the range 
$\gamma$ = 3 -- 10. The
dispersion around the best fit is small as expected since we have removed the
differences due to the orientation angle. 
This result confirms that sources with different kpc scale morphology and
radio power, are similar on the parsec scale. The slope of the correlation
is in agreement with the expected value from equipartition conditions.

The similarity of radio morphology and jet velocity on the parsec scale 
is in contrast with the large difference on the kpc scale between high and low
power radio sources. We briefly discussed the reason of this difference, and
suggested that it could be due to a denser ISM in low power
radio sources. A strong interaction between the relativistic jet and the
surrounding medium can produce a decrease of the jet velocity and therefore
FR I jets will become sub-relativistic after a few hundreds of parsecs
and sub--relativistic within a few kiloparsecs. Conversely, jets 
in FR II sources
are mildly relativistic up to the hot spot region. A systematic higher 
$\gamma$ value in FR II sources is excluded by the present data.

A decrease in the jet velocity in low power sources should give rise to
different morphological properties in parsec scale images. One major difference
could be the presence of a two velocity regime in low power source jets
responsible of the limb-brightened structures. We discussed the 
difficulty
to image such jet structure and we noted that the few sources with a 
limb-brightened structure are low power sources. This could be evidence
 of the interaction 
between the jet and the ISM producing a jet deceleration in low power sources.

\acknowledgements

The authors wish to thank R. Fanti for helpful suggestions and discussions,
and G. Taylor for a critical reading of the manuscript.
We thank the staff of NRAO and EVN telescopes involved in the observations, 
for their help. NRAO is a
facility of the National Science Foundation, operated under cooperative
agreement by Associated Universities, Inc. 
This work was partly supported by the Italian Ministry for University
and Research (MURST) under grant Cofin98-02-32.
GG, LF, and TV acknowledge partial 
financial support from the European Commission TMR Programme, Access to 
Large-Scale Facilities under contract ERBFMGECT950012. This research has 
made use of the
NASA/IPAC Extragalactic Data Base (NED) which is operated by the JPL, 
California Institute of Technology, under contract with the Natonal 
Aeronautics and Space Administration.
\vfill\eject

\appendix

\section{Observations and Data Reduction}

The observational details for each source are given in Table 5:
source names (Col. 1 and 2); observing array (Col. 3); observing
frequency, time (Col. 4 and 5) and observing date (Col. 6).
The observations of the sources 0222+36 (on 97--02--22) and 3C 272.1
were correlated in Bonn, all other observations have been 
correlated in Socorro (NM).

\begin{deluxetable}{lclccc}
\tabletypesize{\small} 
\tablecaption{The new observations\label{tab5}}
\tablehead{
\colhead{Name}&\colhead{Name}&\colhead{Stations}&\colhead{Frequency}&
\colhead{Time}&\colhead{Date} \\
\colhead{IAU}&\colhead{other}&\colhead{ }&\colhead{GHz}&\colhead{hours}&
\colhead{yy-mm-dd}}
\startdata
0116+31  & 4C31.04 & VLBA Y1                       & 5.0 & 4   & 95-07-22 \\
0220+43  & 3C66B   & VLBA Y27 GB EB JB MC NT ON    & 5.0 & 6   & 93-09-12 \\ 
0222+36  &         & VLBA Y1                       & 5.0 & 4   & 95-07-22 \\ 
         &         & EB MC NT ON JB TR             & 5.0 & 2   & 97-02-22 \\
0258+35  & NGC1167 & VLBA Y27 GB EB JB MC NT ON    & 5.0 & 6   & 93-09-12 \\ 
0331+39  & 4C39.12 & VLBA Y1                       & 5.0 & 4   & 95-07-22 \\ 
0648+27  &         & VLBA Y1                       & 5.0 & 4   & 95-07-22 \\ 
1217+29  & NGC4278 & VLBA Y1                       & 5.0 & 4   & 95-07-22 \\
1222+13  & 3C272.1 & BR FD HN LA PT SC JB MC NT    & 1.7 & 10  & 96-02-10 \\ 
1322+36  & NGC5141 & VLBA Y27 GB EB JB MC NT ON    & 5.0 & 6   & 93-09-12 \\ 
1441+52  &  3C303  & VLBA Y1 EB CM MC NT ON TR     & 5.0 & 4   & 97-09-23 \\ 
1833+32  & 3C382   & VLBA Y1 EB CM MC NT ON TR     & 5.0 & 4   & 97-09-23 \\
2243+39  & 3C452.0 & VLBA Y1 EB CM MC NT ON TR     & 5.0 & 4   & 97-09-23 \\
\enddata
\tablecomments{Stations: VLBA = full VLBA (10 telescopes); BR = VLBA-Brewster;
FD = VLBA-Fort Davis; HN = VLBA-Hancock; LA = VLBA-Los Alamos; PT = VLBA-Pie
Town; SC = VLBA-Saint Croix; NL = VLBA-North Liberty; Y27 = VLA phased 
array; Y1 = VLA single antenna; GB = Green Bank (NRAO); EB = Effelsberg; 
CM = Cambridge; MC = Medicina;
JB = Jodrell Bank; NT = Noto; ON = Onsala; TR = Torun; WS = Westerbork
}
\end{deluxetable}

Post correlation processing used the NRAO AIPS package. Amplitude
calibration was initially done using standard method employing measured
system temperatures and assuming sensitivity calibration and refined
using compact calibrator sources. Phase calibration
was done using strong calibrator sources observed during the scheduled time.
All data were globally fringe fitted \citep{sc83} and then self--calibrated.
No polarization information is available.

\section{Notes on Individual Sources} 

{\bf 0055+30 -- NGC315} In previous papers (\citet{cot99} and ref. 
therein) we published 
multi--frequency and multi--epoch data on this giant radio galaxy. We found 
evidence of an acceleration in the inner parsec scale jet. 
A similar result was presented by \citet{lob99}
for the well known quasar 3C 345, by \citet{su00} for NGC 6251,
by \citet{ho00} in the jet of 5 Blazers from proper motion observations
and here
for the source 0220+43 (3C66B).
\citet{bar99} reported the detection in NGC 315 of a polarized broad
H$\alpha$ emission.

{\bf 0104+32 -- 3C31} Observations at 6 cm were published in \citet{lar97}.

{\bf 0116+31 -- 4C31.04} In a previous paper \citet{cot95}
suggested that this source could be a low redshift Compact Symmetric Object
(CSO). \citet{con96} showed the presence of a complex HI absorption across 
both lobes.

We observed this source with the VLBA and one single VLA telescope for 4 
hours at 5 GHz on July 1995. 
In Fig. 5 we present an image obtained with these new data. A comparison with the images
published by \citet{cot95}, confirms the reality of the faint component
in between the two extended lobes and its identification as the core source.
Its spectrum is flat, its flux density being 14 mJy at 8.4 GHz and 14.5 mJy at 
5 GHz. The
spectral index of the East lobe is $\alpha^{1.7}_5$ = 0.5 and 
$\alpha^{5}_{8.4}$ = 0.8; in the W lobe is
$\alpha^{1.7}_5$ = 0.5 and $\alpha^{5}_{8.4}$ = 1.5 (we note however that we
could have missing flux problems at 8.4 GHz as discussed in \citet{cot95}).
This result confirms the identification of 0116+31 (4C31.04) as a nearby CSO source.

No jet-like structure is visible on either side of the core even if both lobes show
an elongated emission in between the region with the highest brightness in the
lobes and the core.
The symmetric structure and the core to total radio power ratio suggests that 
this source is near to the plane of the sky ($\theta$ \gtsim 75$^\circ$). 
This result is in agreement with the \citet{con96} model and explains the lack 
of visible jets, since at this orientation relativistic jets are strongly
de-boosted. 

\begin{figure}
\epsscale{0.8}
\plotone{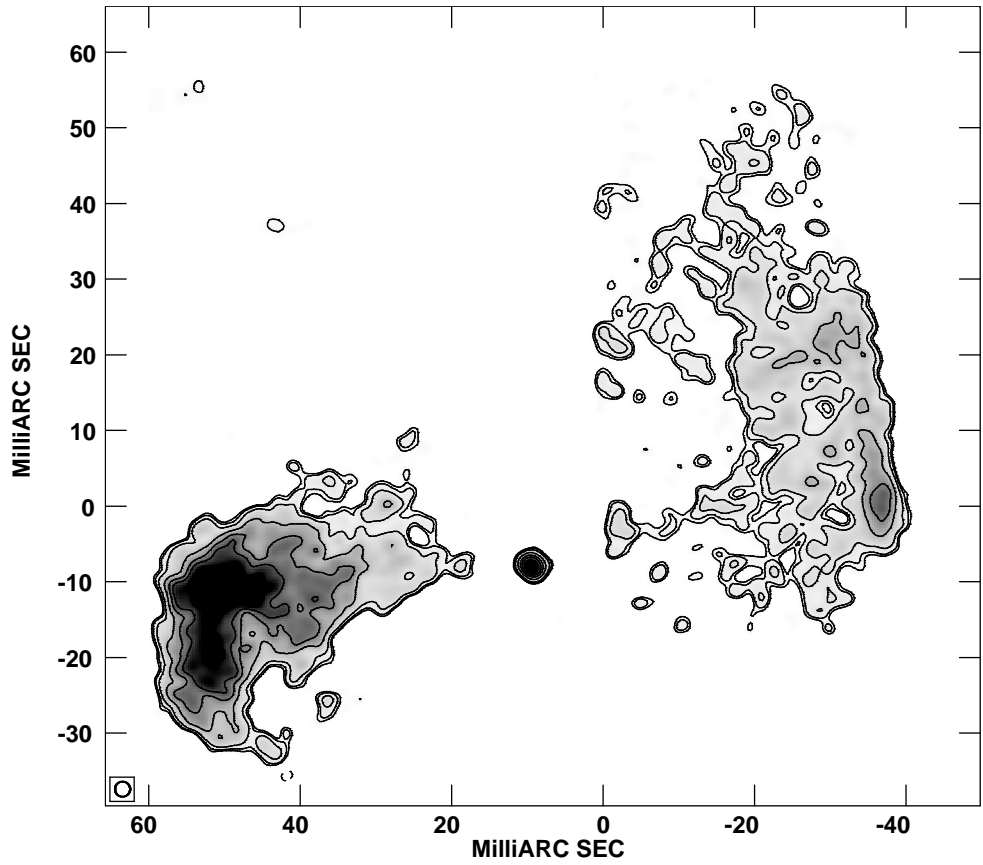}
\caption{VLBA image at 5 GHz of 0116+31 (4C31.04). The HPBW is 2 mas. The noise
level is 0.2 mJy/beam and levels are: -1, 0.8, 1, 1.5, 3, 7, 7, 10, 15
and 20 mJy/beam. The peak flux density is 20.6 mJy/beam.}
\label{f5eps}
\end{figure}

{\bf 0206+35 -- 4C35.03} Observations at 5 GHz have been published in
\citet{lar97}.

{\bf 0220+43 -- 3C66B} This radio galaxy is characterized by the presence of 
an optical and infrared jet (see \citet{jac93} and \citet{tan00}).
At arc-second resolution it was studied in detail by \citet{har96}. 
These authors report also evidence of radio nuclear flux
density variability and studied the kpc scale jet asymmetry. 
A possible optical emission from the counter-jet was reported by \citet{fb97}. 

We observed this source with global VLBI at 6 cm.
The uniform weight map is presented in Figure 6.
The image shows the main jet emission visible up to 20-25 mas from the core.
At higher resolution, a faint counter-jet is visible (Fig. 7). We checked 
accurately
the reality of this faint counter-jet emission, and unsuccessfully tried  
to eliminate it from the image. 
Thanks to the very good uv coverage of our data we are
confident that this structure is real. 
The jet/counter-jet ratio at 1.5 - 2 mas from the core is $\sim$ 10
which corresponds to $\beta$ cos$\theta$ $\sim$ 0.43. 
At larger distance ($\sim$ 5 mas) the jet/counter-jet ratio increases up to R \gtsim 
100 implying a small angle with respect to the line of sight. Taking into 
account the core dominance we derive a jet orientation $\theta \sim 45^\circ$
and a jet velocity $\beta \sim$ 0.6 at 1.5 mas from the core and $\sim$ 0.99
at 4-5 mas from the core. 
This result is very similar to that obtained by us for NGC 315 and by 
\citet{su00} for NGC 6251 and in agreement with \citet{lob99} for 3C345. 

A high jet speed could explain the presence of the gap between the mas and the
arc-second scale, a high velocity jet being de-boosted.  
In the region of the bright optical jet, we need
$\gamma \le$ 2, to avoid de-boosting effects, in agreement with \citet{tan00}.

\begin{figure}
\epsscale{0.5}
\plotone{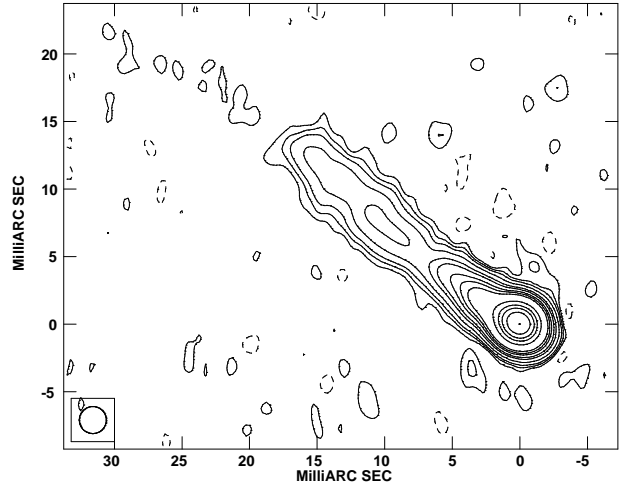}
\caption{Global VLBI image of 0220+43 (3C66B) at 5 GHz. 
The HPBW is 2 mas. The noise
level is 0.06 mJy/beam and levels are: -0.15, 0.15, 0.3, 0.5, 1, 2, 3,
5, 7, 10, 30, 50, 70, 100 and 150 mJy/beam.}
\label{f6eps}
\end{figure}

\begin{figure}
\epsscale{0.5}
\plotone{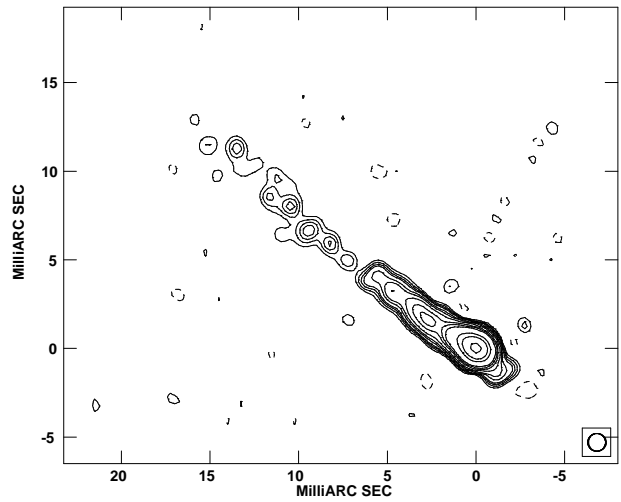}
\caption{As Figure 6 but at full resolution: HPBW = 1 mas. The noise level
is 0.1 mJy/beam and levels are: -0.4, 0.4, 0.6, 0.8, 1, 1.5, 2, 3, 5,
7, 10, 30, 50 and 100 mJy/beam.}
\label{f7eps}
\end{figure}

We note that these values are in agreement with the result $\theta$ \ltsim 
53$^\circ$ estimated by \citet{har96} on the kpc scale.

{\bf 0222+36} This source shows 
a halo-core structure at arc-second resolution
\citep{fan87}. The spectral index between 408 and 5000 MHz is straight
and moderately flat ($\alpha_{0.4}^{5.0}$ = 0.4). We observed this galaxy with the EVN array at
5 GHz on February 1997 and with the VLBA on July 1995. In both observations, 
we detected an unresolved  source with a total flux density comparable to the 
arc-second core flux density. 
From the core to total power ratio we derive that this source should be
oriented at an angle to the line of sight smaller than 40$^\circ$.

{\bf 0258+35 -- NGC1167} This source was studied with VLA and MERLIN+EVN 
observations at 1.6 GHz by \citet{san95} who classified it as a 
CSS source. VLA data show
a double structure with a separation of 1.1''. In the EVN+MERLIN images it
reveals an
extended plume-like feature at both ends of the source and a jet-like feature
in between. Due to the complex structure the core identification
is not obvious. 

We observed this source with global VLBI observations
at 5 GHz. The source is marginally detected at the longest 
baselines but it shows a large increase in the correlated flux density on the 
shortest baselines as EB-JB 
(in the EVN) and Y27-PT-LA (in the US). Our final image shows 
a core-jet structure (Fig. 8), however the large difference between this high
resolution map and the MERLIN+EVN image at 1.6 GHz by \citet{san95}, does not
allow us to properly identify the nuclear source, even if our extension is in the
same direction of the jet-like feature visible in the EVN+MERLIN image.
The peak flux in our image is 26.3 mJy/beam and the total flux density is 34
mJy. We note that in our shortest baseline (Y-PT) a correlated flux of 
$\sim$ 170 mJy is visible confirming the diffuse morphology of the extended 
structure visible in the MERLIN images and its steep spectrum.

\begin{figure}
\epsscale{0.3}
\plotone{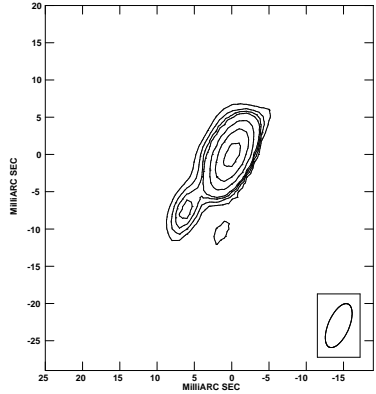}
\caption{Global VLBI image of 0258+35 (NGC1167) at 5 GHz. 
The HPBW is 6.3 $\times$ 2.8
mas (PA = -23$^\circ$). The noise level is 0.2 mJy/beam and levels are:
-1, 1, 1.5, 2, 2.5, 5, 10 and 20 mJy/beam.}
\label{f8eps}
\end{figure}

More observations with shorter baselines are necessary to properly study
this source.
 
{\bf 0331+39 -- 4C39.12} We observed this source with the VLBA and one single 
VLA telescope for 4 hours at 5 GHz on July 1995. The source shows an one-sided 
structure at
the same PA as the short extension visible in the VLA map at 5 GHz 
\citep{fan86}. The parsec scale jet is straight in the inner 12 mas
after which it becomes transversely resolved with many substructures possibly
due to a helical structure although we cannot exclude a limb-brightened
structure (Figure 9 and 10).

\begin{figure}
\epsscale{0.3}
\plotone{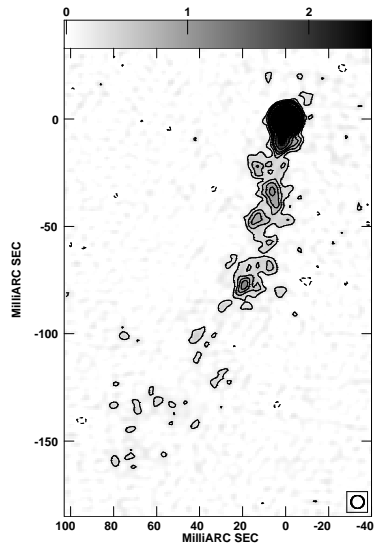}
\caption{VLBA image of 0331+39 with natural weight at 5 GHz. 
The HPBW is 6 mas.
The noise level is 0.08 mJy/beam and levels are: -0.25, 0.25, 0.5, 0.75,
1, 1.5, 3, 5, 10, 20, 30, 50, 70 and 100 mJy/beam.}
\label{f9eps}
\end{figure}

\begin{figure}
\epsscale{0.2}
\plotone{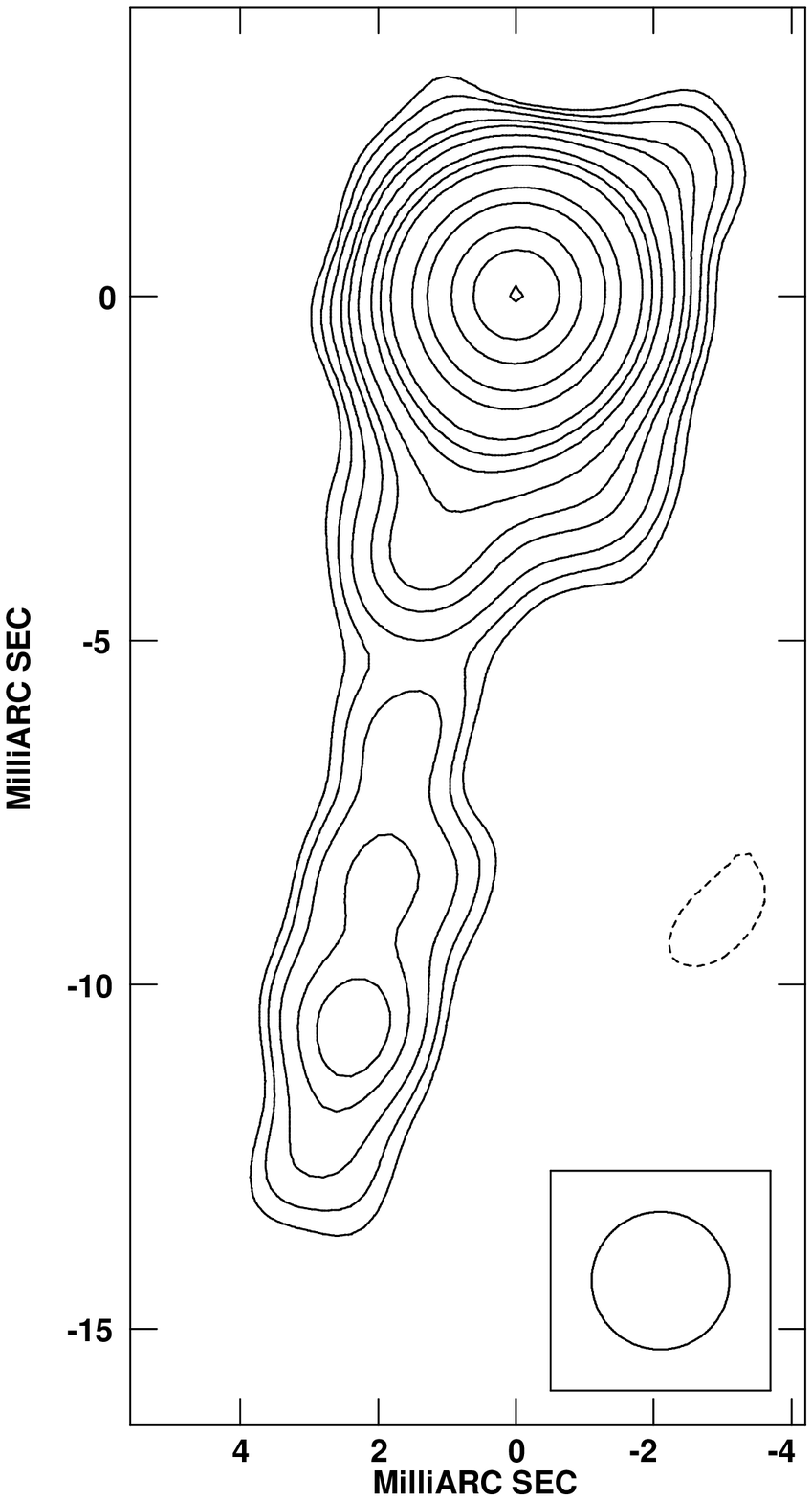}
\caption{VLBA image of 0331+39 at 5 GHz with uniform weight: 
HPBW = 2 mas. The noise level
is 0.15 mJy/beam and levels are: -0.5, 0.5, 0.7, 1, 1.5, 2, 3, 5, 7,
10, 20, 30, 50, 70 and 90 mJy/beam.}
\label{f10eps}
\end{figure}

The limit on the jet/counter jet ratio is not very high because of the 
jet low brightness
(R \gtsim 12) therefore the major constraints on the jet orientation and 
velocity
were derived from the core dominance, which gives  $\theta$ lower than
45$^\circ$.
A small angle with respect to the line of sight is in agreement
with the halo-core structure of the radio emission at arc-second resolution.

{\bf 3C109} We published VLBI and VLA observations of this FR II Broad Line
Radio Galaxy (BLRG) in \citet{g94}.

{\bf 0648+27} This source is unresolved at arc-second resolution and has a 
spectral index $\alpha^{5.0}_{0.4}$ = 0.6 suggesting the existence of a radio
structure in the mas or sub-arcsecond scale. The measured flux densities at
arc-second resolution are:
S$_{408}$ 270 mJy, S$_{1415}$ 156 mJy 
\citep{fan87} and S$_{5000}$ 58 mJy \citep{gre91}, while
\citet{ant85} reports a flux density of 213 mJy with VLA B array 
observations at 5 GHz. 

We observed this galaxy with VLBA + Y1 at 5 GHz on July 1995 
and found an unresolved core with a flux density of 47 mJy and a faint
one-sided emission extended 20 mas with a variable direction
(Fig. 11).
From this asymmetric structure and the core dominance we derive that this
source is oriented at $\theta$ \ltsim 40$^\circ$ with a jet velocity
\gtsim 0.7c.

\begin{figure}
\epsscale{0.4}
\plotone{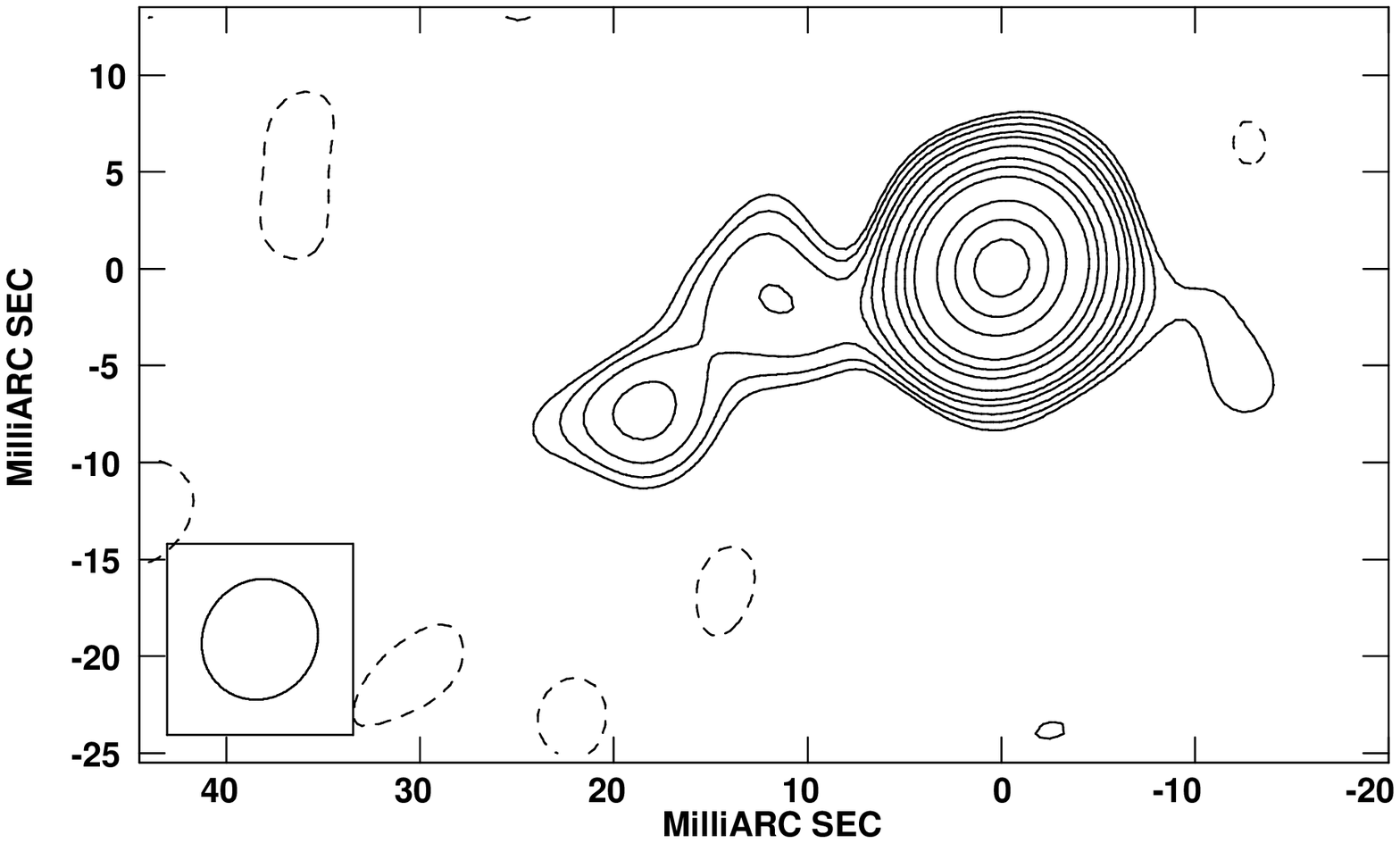}
\caption{VLBA image of 0648+27 at 5 GHz. The HPBW is 6.4 
$\times$ 5.9 mas (PA -30$^\circ$).
The noise level is 0.2 mJy/beam and levels are: -0.5, 0.5, 0.7, 1,
1.5, 2, 3, 5, 7, 10, 20, 30 and 40 mJy/beam.}
\label{f11eps}
\end{figure}

{\bf 0755+37 -- NGC2484}  We published VLBI and VLA observations of this FR I
Radio Galaxy in \citet{g94}.

{\bf 0836+29 -- 4C29.30} We published VLBI and VLA observations of this FR I
Radio Galaxy obtained on November 1990, in \citet{ven95}. A second epoch
observation always at 6 cm was obtained on May 1994. A comparison of the two
images does not reveal any proper motion. 
More efforts, as in NGC 315, are necessary to measure a possible proper motion
in this source. 

{\bf 1101+38 -- Mkn 421} This BL-Lac source was studied in detail by 
\citet{pin99}. We presented a VLBA image in \citet{g99b} and a more 
detailed paper is in preparation. The core dominance for this source is
low (to be a BL-Lac) even if we take into account the core variability
\citep{pin99}. 
The presence of proper motion is uncertain:
from our data we derive a possible proper motion of $\sim$ 1.5 c (Giovannini et
al., in preparation), while 
\citet{pin99}
found sub luminal proper motions (or no motion at all) and \citet{mar99} found
a proper motion with an apparent velocity of the order 2-3 c.

Taking into account the jet/counterjet ratio and the core dominance,
we can have: a) a high velocity jet but in this case the angle with respect 
to the line of sight should be near 20$^\circ$; b) a jet oriented at a small
angle ($\sim 5^\circ$) with $\beta$ as low as 0.9.
In either cases we have a low Doppler factor. If we want to reconcile this
result with the high Doppler factor requested by gamma ray and variable 
emission,
we need a) a large change in the jet orientation from the TEV to
the radio emission region; or b) a large decrease of the jet velocity
from the TEV to the radio emission region. Given the radio morphology
of the pc scale jet with its diffuse emission, both cases (a) and (b)
are probably present.

{\bf 1142+20 - 3C264} We discussed the properties of this source in 
\citet{lar97}, \citet{bau97}, \citet{lar99} and Lara et al., in preparation.
We assume here as jet dynamics the results given in \citet{bau97}.
This source is classified as a low luminosity BL Lac by \citet{rec99}.

{\bf 1144+35} See \citet{g99a} for a detailed discussion on this 
superluminal source.

{\bf 1217+29 -- NGC 4278} This nearby (z = 0.0021; distance modulus = 30.61,
corresponding to 13.2 Mpc) elliptical galaxy shows strong nuclear optical
emission lines and a large amount of neutral hydrogen (see \citet{sch83}
and references therein).
At arc-second resolution it shows an unresolved emission, while an extended 
structure is visible at mas resolution \citep{sch83}.

\begin{figure}
\epsscale{0.6}
\plotone{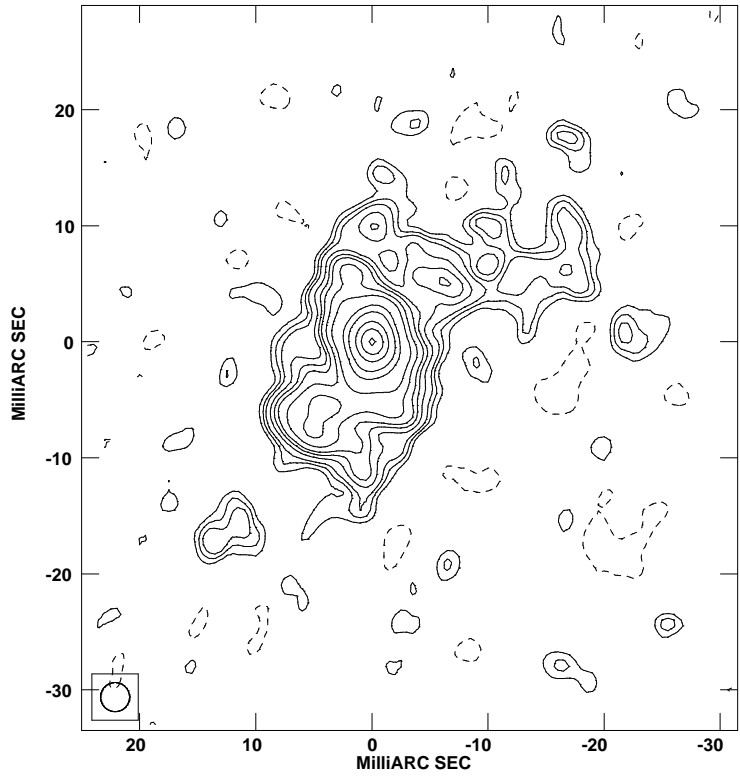}
\caption{VLBA image of 1217+29 (NGC4278) at 5 GHz. The HPBW is 
2.5 mas. The noise level
is 0.4 mJy/beam and levels are: -1, 1, 1.5, 2, 3, 4, 5, 7, 10, 20, 30,
50, 70, and 90 mJy/beam.}
\label{f12eps}
\end{figure}

We observed this source with VLBA + Y1 on July 1995 at 5 GHz. Our image
(see Figure 12) shows a complex structure with a central emission with a peak
flux density of 95 mJy/beam and an extended halo which at North shows a
large bend, being oriented E-W in the external regions. 
A distorted jet-like structure could be present in the southern region,
however due to the extended structure also in the northern
region, no information on the jet velocity and
orientation may be derived from present data.
Our image is in good agreement with the map presented by 
\citet{sch83} at a lower angular resolution and with the image given by 
\citet{fal00}
where only the southern jet like structure is visible because of their lower
sensitivity to extended structures (they
do not have the short baseline VLA - VLBA Pt). A comparison with unpublished
images at 8.4 GHz by us, confirms the core identification with the 
central region
where the highest brightness is measured.
The VLBI total flux density is $\sim$ 400 mJy in agreement with arc-second scale
observations, therefore we are confident to have mapped the whole structure
of this small size source.
We note that due to the nearness of this source
the linear resolution of our image is very good (the HPBW is 2.5 mas 
corresponding to 0.15 pc).

{\bf 1222+13 -- 3C272.1} A preliminary map of this source at 5 GHz presented 
in \citet{g95}
suggested a possible two-sided structure. We present now a new
better image obtained at 1.7 GHz
where this nearby source shows a clear one-sided
structure (Fig. 13). A new reduction of previous 5 GHz data confirms the 
one-sidedness of this source. The main problem in the 5 GHz data is the poor 
uv-coverage with large 
gaps and the relatively faintness of the extended emission. 
The present image is in agreement with results obtained in different 
observations using the full VLBA by \citet{xu00}.

\begin{figure}
\epsscale{0.25}
\plotone{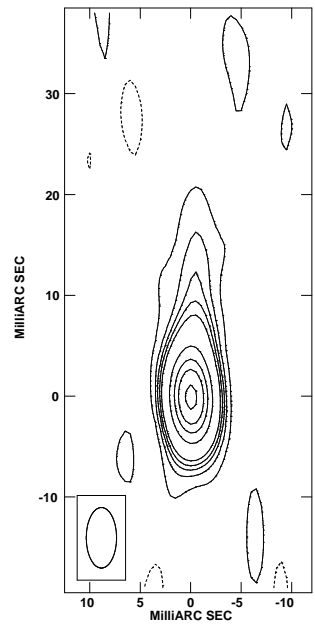}
\caption{Global VLBI image of 1222+13 (3C272.1) at 1.7 GHz. The HPBW is 6 $\times$ 3
mas in PA 0$^\circ$. The noise level is 0.5 mJy/beam and levels are:
-1, 1, 3, 5, 7, 10, 30, 50, 70, and 100 mJy/beam.}
\label{f13eps}
\end{figure}

From the present data, we derived a limit on the ratio between the jet and 
counter-jet \gtsim 10 which
implies $\theta$ \ltsim 65$^\circ$. The low core radio 
power suggests that it is strongly de-boosted and 
therefore at a large angle with respect to the line of sight.
We derive that $\theta$ has to be \gtsim 60$^\circ$.
Taking into account both constraints we estimate $\theta$ $\sim$ 60$^\circ$ 
- 65$^\circ$ 
with $\beta$ \gtsim 0.9. This result is in agreement with the 
asymmetry visible at the beginning of the arc-second jet (see the unpublished
image by Laing and Bridle in {\it An Atlas of DRAGNs} - 
http://www.jb.man.ac.uk/atlas/) and with the gap of radio emission in the 
sub-arcsecond scale.

{\bf 1228+12 -- 3C274} This source is well studied at all frequencies and
angular resolutions. See e.g. \citet{jun99}, \citet{bir99} and references in.

{\bf 1322+36 -- NGC5141} We observed this small size FR I radio galaxy 
for 6 hours at 5 GHz on September 1993.
The final image shows an one-sided jet emission extended $\sim$ 10 mas in 
the same direction of the main large scale jet (Fig. 14).

\begin{figure}
\epsscale{0.4}
\plotone{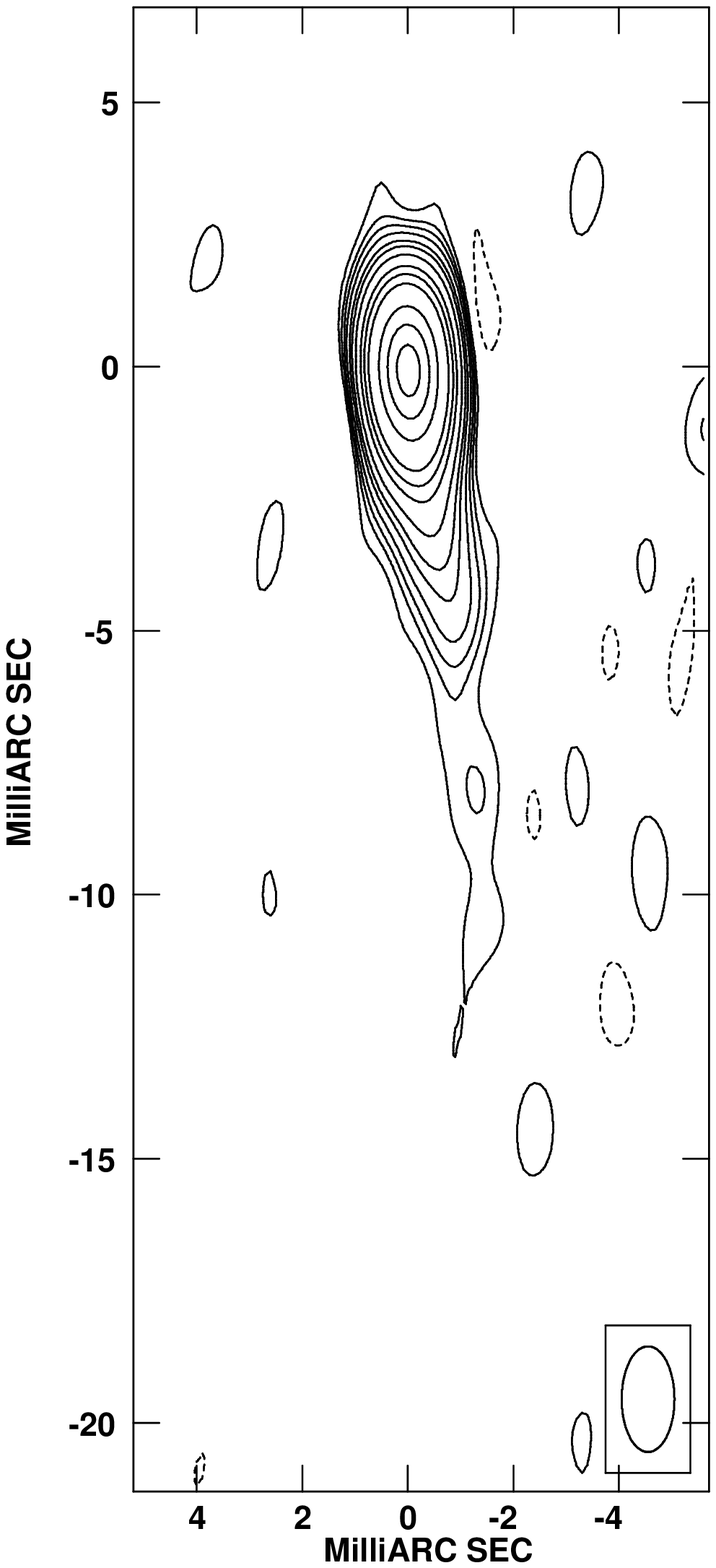}
\caption{Global VLBI image of 1322+36 (NGC5141) at 5 GHz. The 
HPBW is 2 $\times$ 1 mas in
PA 0$^\circ$. The noise level is 0.14 mJy/beam and levels are: -0.3, 0.3, 0.5, 0.7, 
1, 1.5, 2, 3, 5, 7, 10, 20, 30, and 40 mJy/beam.}
\label{f14eps}
\end{figure}

We derive a jet/counter-jet ratio $>$ 20 which gives a value of $\beta cos\theta >$
0.54 implying a jet velocity $\beta$ \gtsim 0.54 and an orientation angle
$\theta$ \ltsim 58$^\circ$. 

{\bf 1441+52 -- 3C303} This Broad Line Radio Galaxy shows an extended 
structure at arc-second 
resolution, two hot spots and a prominent one-sided jet
\citep{lea91}. We have observed this source with a
global array of 17 telescopes for 6 hrs at 5 GHz. Our final map
is in Fig. 15. The parsec scale image shows an one-sided jet in the same
orientation of the arc-second scale jet. In the inner 5 mas, the radio emission
is not resolved transversely while in the 5 - 15 mas region the jet emission
becomes fainter and oscillating.

\begin{figure}
\epsscale{0.4}
\plotone{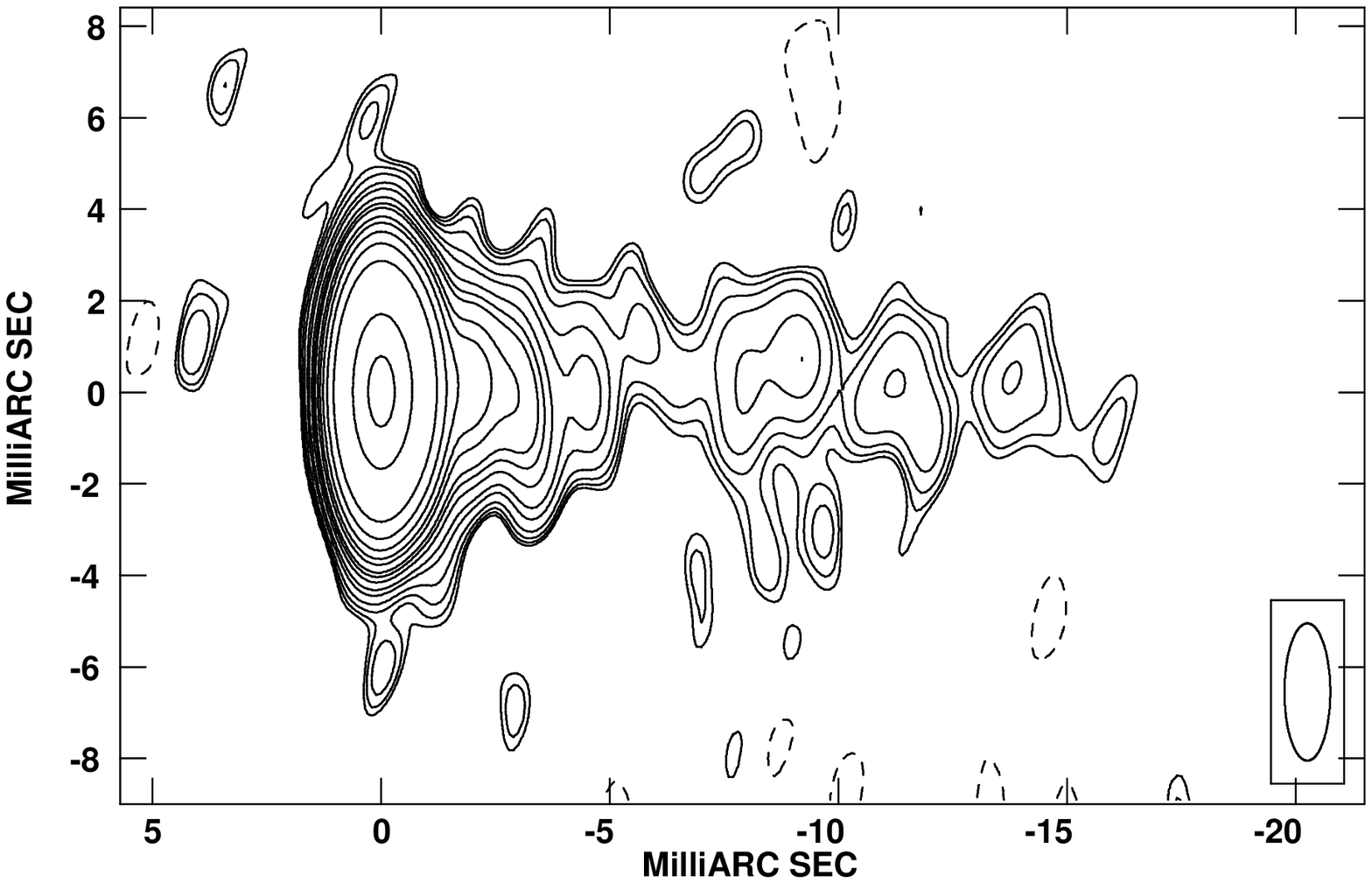}
\caption{Global VLBI image of 1441+52 (3C303) at 5 GHz. The HPBW 
is 3 $\times$ 1 mas in
PA 0$^\circ$. The noise level is 0.05 mJy/beam and levels are: -0.1, 0.1, 0.12, 
0.15, 0.2, 0.3, 0.4, 0.6, 0.8, 1, 1.5, 2, 3, 5, 10, 50, and 100 mJy/beam.}
\label{f15eps}
\end{figure}

From the jet sidedness and the core prominence we derive that the radio emission
is oriented at an angle $\theta$ \ltsim 40$^\circ$ and the jet velocity has to
be \gtsim 0.7c. 
The derived jet orientation and velocity is in agreement with the presence of 
broad lines in the optical spectrum and with the arc-second radio structure.

{\bf 1626+39 -- 3C338} We are monitoring this source because of its two-sided
structure and measured proper motion. Our last paper on this source is
\cite{g98a}.

{\bf 1641+17 -- 3C346} See \cite{cot95} for a discussion on the parsec
scale structure of this Broad Line Radio Galaxy.

{\bf 1652+39 -- Mkn 501} We observed this source also with the space VLBI 
array. Preliminary results have been published in \cite{g99b} and \cite{g98b}.
A more detailed paper is in preparation.  

{\bf 1833+32 -- 3C382} An 8.4 GHz VLBI image of this source was published
in \cite{g94}.  We present here new observations with a
global array of 17 telescopes for 6 hrs at 5 GHz. 
The new image is in good agreement with the old one. Thanks
to the better sensitivity and uv coverage we see a more extended jet with a
peak brightness at its end (Fig. 16). The possible change in the jet PA near the 
core is not confirmed by the new data. A comparison of the 8.4 and 5 GHz images
shows some
displacement between local peaks possibly due to proper motion, but the
large epoch difference and the very different uv coverage (and frequency)
does not allow us to estimate it.

\begin{figure}
\epsscale{0.4}
\plotone{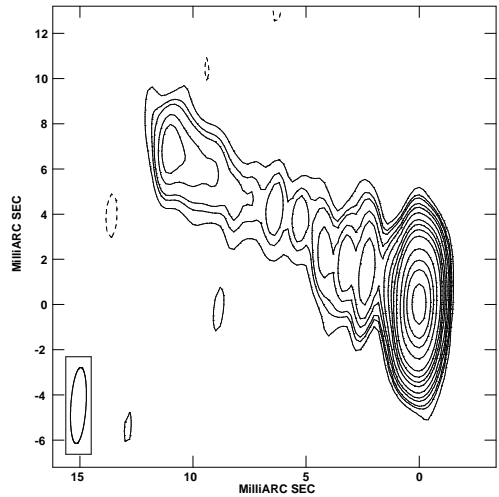}
\caption{Global VLBI image of 1833+32 (3C382) at 5 GHz. The HPBW 
is 3.4 $\times$ 0.7 mas
in PA = -4$^\circ$. The noise level is 0.15 mJy/beam and levels are: -0.5, 0.5, 
0.8, 1, 1.5, 2, 3, 5, 7, 10, 15, 30, 50, 70, and 100 mJy/beam.}
\label{f16eps}
\end{figure}

\begin{figure}
\epsscale{0.7}
\plotone{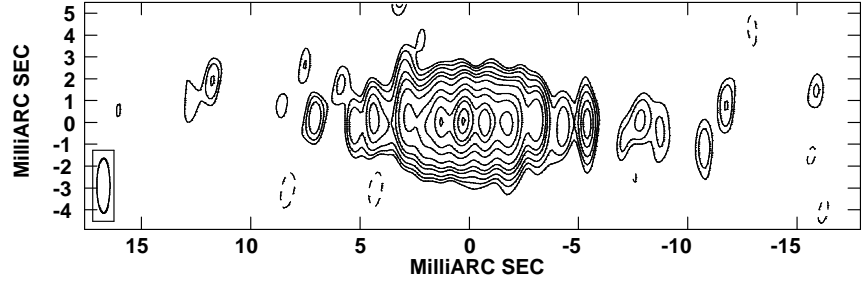}
\caption{Global VLBI image of 2243+39 (3C452) at 5 GHz, uniform 
weight. The HPBW is
2.5 $\times$ 0.6 mas in PA 0$^\circ$. The noise level is 0.2 mJy/beam and levels are:
-0.5, 0.5, 0.7, 1, 1.5, 2, 3, 5, 7, 10, 12, 15, and 17 mJy/beam.}
\label{f17eps}
\end{figure}

\begin{figure}
\epsscale{0.6}
\plotone{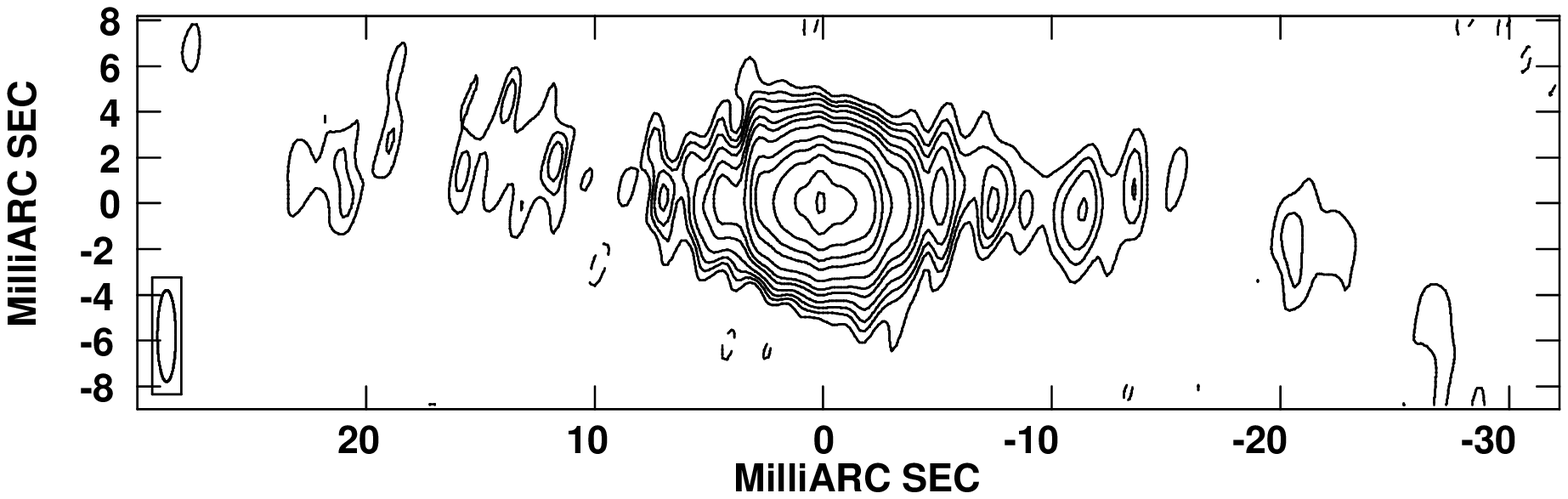}
\caption{Global VLBI image of 2243+39 (3C452) at 5 GHz, natural 
weight. The HPBW is 
4 $\times$ 0.8 mas in PA 0$^\circ$. The noise level is 0.1 mJy/beam and levels are:
-0.3, 0.3, 0.5, 0.7, 1, 1.5, 2, 3, 5, 7, 10, 15, and 20 mJy/beam.}
\label{f18eps}
\end{figure}

{\bf 1845+79 -- 3C390.3} This super-luminal Broad Line Radio Galaxy has been 
studied by \cite{al96}. They suggest a proper motion with an apparent
velocity of 3.5 c. Taking into account this result and the core dominance
(the main jet is faint, therefore no strong constraint can be derived from
the jet sidedness) we derive that $\theta$ is in the range 30-35 degree with
$\beta$ $\sim$ 0.96 - 0.99. The derived low Doppler factor is in agreement
with the faint pc scale jet. The value of $\theta$ is in agreement with the
kpc scale structure and the presence of broad lines in the nuclear region.

{\bf 2243+39 -- 3C452} We have observed this FR II narrow line and symmetric
radio galaxy with a global array of 17 telescopes for 6 hrs at 5 GHz.
The parsec scale images show a very symmetric structure (Fig. 17). We identify 
the core
source as the central peak taking into account also the symmetric kpc scale
structure, but new observations at different frequencies
are necessary to confirm this result. The pc scale jet position angle 
(90$^\circ$) in the inner 5 mas from the core is 
slightly different from the kpc scale one ($\sim$ 75$^\circ$).
However in a lower resolution image (Fig. 18) a low surface brightness is visible
up to 20 mas from the core possibly oriented as the kpc scale structure.
The pc scale jet brightness shows two different regions: it is high up to 
$\sim$ 5 mas from the core and shows a strong decrease at larger distances.
From the jet symmetry and the core dominance we derive that this source
should be oriented at $\theta$ \gtsim 60$^\circ$. 

{\bf 2335+26 -- 3C465} This extended WAT source was studied in \cite{ven95}.
A second epoch global observation obtained on September 1993 at 8.4 GHz 
(not shown here)
does not reveal any visible proper motion, because of the short time range and
of the uniform brightness of the parsec scale jet.
More effort is necessary to measure a possible proper motion
in this source.   

\vfill\eject

\end{document}